\documentclass[11pt]{article}
\usepackage[margin=1.0in]{geometry}

\usepackage{comment}

\usepackage{graphicx}
\graphicspath{{./images/}}

\usepackage{multicol}        

\usepackage{xcolor}

\usepackage{bbm}

\usepackage{xspace}
\newcommand{\MAP}{{\ensuremath{MAP}\xspace}}

\usepackage{tikz}
\usetikzlibrary{shapes,arrows.meta,calc,positioning}
\usepackage{subcaption}
\usepackage{hyperref}
\usepackage{amsmath,amssymb,amsfonts}

\begin{document}

\title{
A flow-kick model of dryland vegetation patterns: the impact of rainfall variability on resilience
}

\author{Punit Gandhi\thanks{Department of Mathematics and Applied Mathematics, Virginia Commonwealth University, Richmond, VA 23284, USA  (\href{mailto:gandhipr@vcu.edu}{gandhipr@vcu.edu})}
	\and Matthew Oline\thanks{
Computational and Applied Mathematics,  University of Chicago, Chicago, IL 60637, USA}
	\and Mary Silber\thanks{Department of Statistics and Committee on Computational and Applied Mathematics,  University of Chicago, Chicago, IL 60637, USA;
    NSF-Simons National Institute for Theory and Mathematics in Biology
}
	}

\maketitle
	
\begin{abstract}
In many drylands around the globe, vegetation self-organizes into regular spatial patterns in response to aridity stress. We consider the regularly-spaced vegetation bands, on gentle hill-slopes, that survive low rainfall conditions by harvesting additional stormwater from upslope low-infiltration bare zones. We are interested in the robustness of this pattern formation survival mechanism to changes in rainfall variability. For this, we use a flow-kick modeling framework that treats storms as instantaneous kicks to the soil water. The positive feedbacks in the storm-level hydrology, that act to concentrate water within the vegetation bands, are captured through the spatial profiles of the soil water kicks. Between storms, the soil water and vegetation, modeled by a two-component reaction-diffusion system, evolve together.  We use a combination of linear stability analysis and numerical simulation to compare predictions of idealized periodic rainfall, with no variability, to predictions when there is randomness in the timing and magnitude of water input from storms. We show that including these random elements leads to a decrease in the  parameter range over which patterns appear. This suggests that an increase in storm variability, even with the same mean annual rainfall, may negatively impact the resilience of these pattern-forming dryland ecosystems. \vspace{2mm}\\

\noindent\textbf{keywords:} pattern formation, dryland ecosystems, vegetation patterns, flow-kick dynamics, random dynamical systems, impulsive reaction diffusion systems.	

    \end{abstract}



\section{Introduction}
\label{sec:intro}

Vegetation in dryland environments must incorporate strategies for surviving their water-limited conditions~\cite{maestre2021biogeography,maestre2016structure}. A spectacular one of these, which acts at the community scale, is spontaneous self-organization into regularly-spaced patches that can harvest water from the surrounding area following a rare rainstorm~\cite{menaut2001banded,deblauwe2008global,gandhi2019vegetation}. Early observational studies of large-scale banded vegetation patterns in the Horn of Africa provided some insights on the phenomena~\cite{macfadyen1950vegetation,greenwood1957development,hemming1965vegetation}.
These investigations showed how banded vegetation, with bands oriented transverse to gentle slopes, can collect the runoff from upslope bare regions that have lower infiltration rates. Subsequent remote sensing studies found similar banded vegetation patterns, to those in Africa, in drylands of Australia and North America~\cite{deblauwe2008global,deblauwe2012determinants}. These empirical investigations reported on pattern characteristics such as band spacing, band width, gradual up-slope migration speeds, as well as some of the changes to the patterns that have occurred over the past 60+years since the first aerial photographs of them~~\cite{deblauwe2012determinants,penny2013local,gowda2018signatures,bastiaansen2018multistability}. 

\begin{figure}
	
	\includegraphics[width=\textwidth]{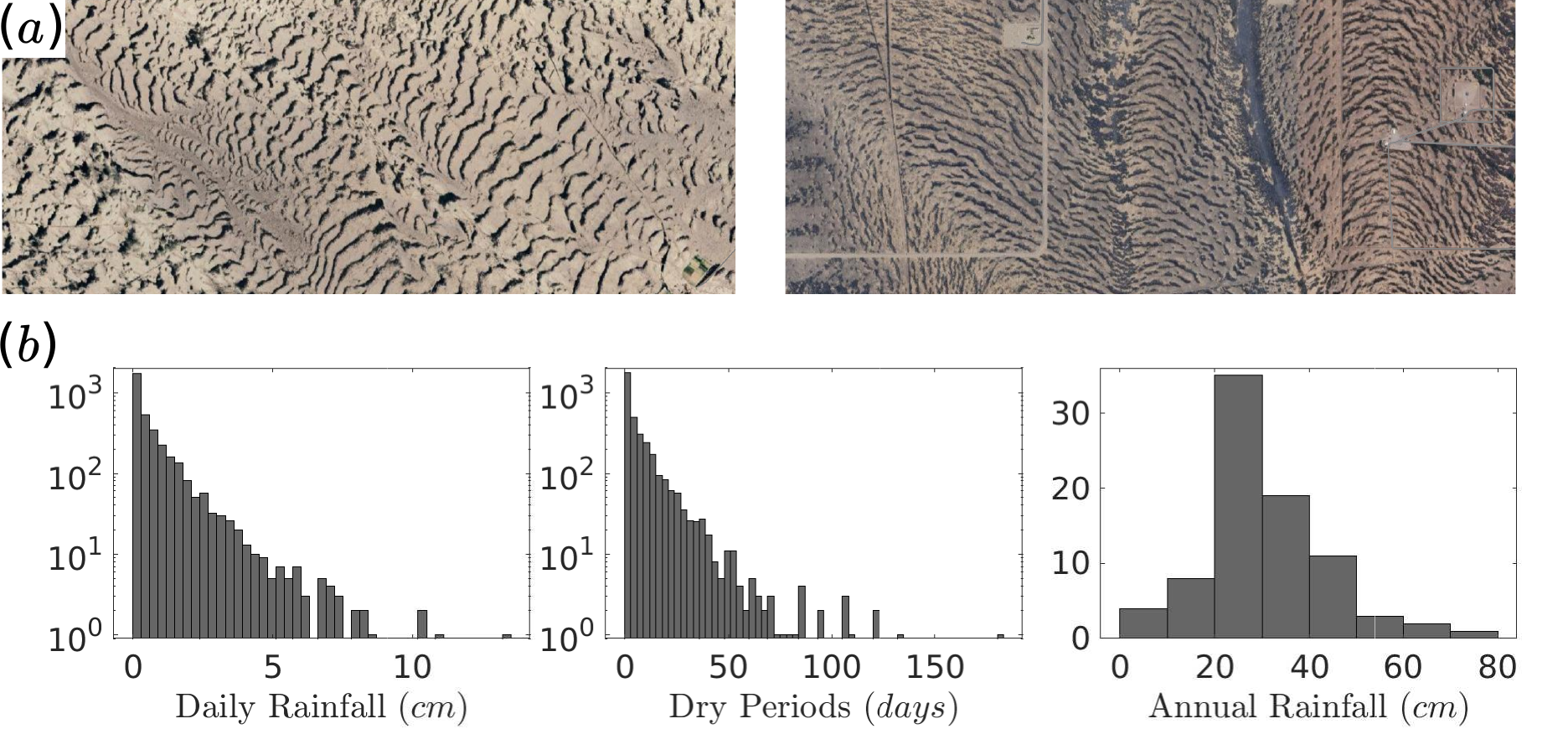}
	\centering
	\caption{
		(a) Satellite images of banded vegetation patterns from sites in Mexico \href{https://www.google.com/maps/@27.93654,-103.3670647,5631m/data=!3m1!1e3?entry=ttu}{(27$^\circ$56'N, 103$^\circ$22'W)} and the USA \href{https://www.google.com/maps/@31.0507499,-103.0904153,3861m/data=!3m1!1e3?entry=ttu}{(31$^\circ$02'N, 103$^\circ$05'W)} taken from Google Maps~\cite{googleMAPS}. 
		(b) Rainfall data from weather station nearby the USA site (Pecos County Airport, Texas, USA) taken between 1940 and 2024~\cite{NOAAcoop}. From left to right, distribution of daily rainfall, dry periods between days with rain, and annual rainfall. 
		\label{fig:intro}} 
\end{figure}

To accompany the empirical studies, idealized mathematical models of vegetation pattern formation have been developed to investigate how changes in aridity might impact pattern characteristics, such as the pattern morphology for 2-d isotropic patterns
or the pattern wavelength for the oriented bands on hillslopes~\cite{Klausmeier1999,lejeune1999short,rietkerk2002self,gilad2004ecosystem}. These modeling and analysis studies have also indicated that pattern-formation mechanisms help the ecosystem to avert the collapse that would otherwise occur if the vegetation was distributed uniformly~\cite{bastiaansen2020effect,rietkerk2004self}. Many of these models take the form of advection-reaction-diffusion PDEs that  capture the interactions between water and biomass on the long ecosystem timescale, with precipitation represented by its annual average value~\cite{gandhi2019vegetation,meron2015nonlinear,martinez2023integrating}.  An alternative approach, which we adopt here, is to treat rain events as impulses to the system, which have the random timing and random strength of a jump process. This approach aligns with Noy-Meir's defining characterization of desert 
environments as  “water-controlled ecosystems with infrequent, discrete, and largely unpredictable water inputs"~\cite{noy1973desert}. He also goes on to emphasize that, while other energy and nutrient inputs to ecosystems may be continuously supplied, the limiting resource of dryland ecosystems arrives in short-lived `pulses'. 
See Figure~\ref{fig:intro} for examples of banded vegetation patterns in North America, along with nearby rainfall data that shows its variability in timing and amount. 

The question we ask in this paper is whether rainfall variability encourages or discourages, compared to an idealized predictable rainfall, the formation of vegetation patterns under conditions of aridity stress.  This is motivated by a larger
question of whether increases in precipitation variability, expected with climate change ~\cite{IPCC_2021,maestre2012getting},  might threaten the effectiveness of vegetation patterns as a mechanism to evade possible ecosystem collapse under a changing climate. We pursue this line of inquiry with a focus on banded vegetation patterns, on gentle slopes, which are modeled on a
one-dimensional spatial domain by the pulsed-precipitation model introduced by Gandhi et al.~\cite{gandhi2023pulsed}. The rainfall arrival times are modeled as a Poisson point process so that the between-storm time intervals are exponentially distributed. We further assume that the depth of storm water, deposited instantaneously and uniformly on the surface, is also an exponentially distributed random variable. The resulting impulse, or `kick', to the soil moisture field from a storm is determined by a simple PDE model for surface water transport and infiltration into the soil, where both down-slope flow speed and infiltration rate depend on the spatial distribution of vegetation. These rates capture the positive feedbacks between biomass and soil moisture, and lead to a heterogeneous increase of soil moisture if vegetation is heterogeneously distributed. During the long dry periods between storms, the soil moisture and biomass fields then evolve, on the slow ecosystem time-scale, according to a simple deterministic reaction-diffusion model. 

The pulsed-precipitation model, as shown in~\cite{gandhi2023pulsed}, predicts that surface hydrology during rainstorms plays a key role in selecting the preferred wavelength of dryland vegetation bands on gentle hillslopes.  Floquet theory, in the case of (temporally) periodic rainfall, determined the (spatial) wavelengths associated with pattern-forming instabilities of the spatially uniform vegetation state. Interestingly, this linear analysis revealed a series of spatial resonance tongues positioned at wavelengths where water from newly-forming bare soil regions can be efficiently harvested by newly-forming vegetation bands. Model simulations indicated that this result carries over to the case of random rainfall -- the mean distance that surface water travels before infiltrating into the soil aligns with the distance from bare soil to the closest  down-slope vegetation band in fully formed patterns.  Here, we develop tools to investigate the instabilities leading to patterns in the case of random rainfall. This allows us to systematically explore the impact of randomness on the pattern-forming instability through linear stability analysis, as well as direct numerical simulation.

There has been extensive research on how pattern formation enhances the resilience of dryland ecosystems, with most of those studies focused on the ecosystem response to decreased mean annual precipitation~\cite{rietkerk2004self,von2001diversity,siteur2014beyond,gowda2016assessing}.  
These investigations suggest that pattern formation occurs as a survival strategy under environmental stress.  
Here we add to that body of research by deploying a modeling framework that allows us to investigate resilience under changes in the variability of the discrete precipitation events that maintain the ecosystem.  One possibility is that rainfall variability increases environmental stress on the ecosystem, and therefore induces earlier onset of pattern formation as mean annual precipitation is decreased.  On the other hand, these  patterns are associated with a response to environmental stress and, therefore, certain characteristics of the emerging patterns (e.g. band spacing and width) may be adapted to survive within the particular environmental conditions in which they form.  This perspective would suggest that increasing rainfall variability, all else being equal, may inhibit the tendency to pattern formation.

We frame this study using some characteristics proposed to quantify an ecosystem's resilience in the face of recurrent `disturbances'. For this,  Meyer et al.~\cite{meyer2018quantifying} advocated for a flow-kick framework, where the flow of an ODE models ecosystem evolution between periodically applied disturbances in the form of kicks to the system state. In their examples, ecosystem collapse occurs on crossing a boundary in the (kick-period, kick-strength)-parameter plane, and a system's proximity to such a boundary is an indication of its resilience to recurrent disturbances. 
We determine a similar boundary for the pattern-forming instability of the uniform vegetation state. 
If the storm kicks become too-infrequent and/or are too-small, then the uniform vegetation state becomes unstable and patterns emerge in response to crossing this boundary. 
We also determine a second boundary, at even worse rainfall conditions, where the bare soil state is stabilized. On crossing this boundary, a catastrophic collapse  of the ecosystem, from the patterned-state to a zero-productivity one, becomes possible. We compute the resilience boundaries for both periodic rainfall and random rainfall and demonstrate that the randomness reduces the regime of resilience of the dryland ecosystem. 

Many early modeling studies of banded vegetation patterns have laid foundations for our investigation. Perhaps the most widely studied, from a mathematical analysis perspective~\cite{sherratt2005analysis,van2013rise,siero2015striped,carter2018traveling}, is the model by  Klausmeier~\cite{Klausmeier1999} that captures,  phenomenologically, feedbacks between biomass and water that lead to pattern formation. It is posed on the ecosystem timescale, as is the case for important spin-offs of it, such as the Rietkerk model~\cite{rietkerk2002self} and the Gilad model~\cite{gilad2004ecosystem,meron2015nonlinear}, which aim to better capture the infiltration feedback by subdividing Klausmeier's water field into surface and soil components. 
These models of vegetation pattern formation work by  adjusting parameters to effectively slow down the fast hydrological processes that convert surface water to soil moisture.  
Other modeling studies have focused on capturing the influence of seasonality and rainfall intermittency on vegetation patterns~\cite{dordorico2006vegetation,ursino2006stability,guttal2007self,konings2011drought,siteur2014will,gandhi2020fast,gordillo2023intermittent}.
Among these, Gandhi et al.~\cite{gandhi2020fast} developed a fast-slow framework for vegetation patterns aimed at using more accurate hydrological parameters. That model was subsequently simplified to become the pulsed precipitation model~\cite{gandhi2023pulsed} that we use in this paper. These models incorporate the runoff-runon mechanism for vegetation pattern formation on hillslopes, which was also incorporated in a particularly  simple way in Siteur et al.~\cite{siteur2014will}, where rainfall was partitioned into a portion that infiltrates the soil and a portion that becomes runoff, with a goal of investigating the impact of changing storm intensity. There are also studies that more accurately model surface water flow, via shallow water equations, in order to investigate vegetation pattern formation~\cite{crompton2021sensitivity}. There a computational speed-up was achieved by deploying a trained statistical emulator for fluid flow over a surface with vegetation patches on it~\cite{crompton2019emulation}. 
Our work is one of a handful of studies that incorporate stochasticity into the model to determine impacts of variability on vegetation pattern formation, which include~\cite{dordorico2006vegetation,konings2011drought,gordillo2023intermittent,kletter2009patterned,yizhaq2014effects,yizhaq2017geodiversity,hamster2024blurring}. 

This paper is organized as follows. In Section~\ref{sec:kickflow}, we introduce the flow-kick model that forms the basis of our study.  Throughout, we compare predictions of this model under random rainfall to those with periodic rainfall in order to specifically assess the impact of rainfall variability.   
In Section~\ref{sec:transcrit}, we focus on the bare soil state of the model, showing that it destabilizes, giving way to a uniformly vegetated state, at slightly higher precipitation levels when randomness is introduced.  In Section~\ref{sec:pattern}, we use Lyaponuv exponents to characterize the linear stability of these uniform vegetation states to pattern forming instabilities under random rainfall, and observe a decrease, compared to periodic rainfall, in the predicted precipitation level for onset of pattern formation.  In Section~\ref{sec:sim}, we numerically confirm the linear predictions through numerical simulation and observe a decrease in the precipitation interval over which the fully nonlinear patterns appear. 
Finally, in Section~\ref{sec:discussion},  we conclude with a discussion of our results in the context of other related studies. This discussion also advocates for incorporating rainfall variability into vegetation pattern models, in light of its potential impact as demonstrated by this investigation, and also because of its likelihood to increase under climate change.  

\section{Flow-Kick Model of Vegetation in Drylands}
\label{sec:kickflow}

The flow-kick model, developed by Gandhi et al.~\cite{gandhi2023pulsed}, captures consumer-resource interactions between biomass density $B(X,T)$ and soil moisture $W(X,T)$. It takes the form of an impulsive reaction-diffusion equation, in which intermittent storms provide jumps $\Delta W_i(X)$ to the soil moisture at times $T_i$. These jumps, or ``kicks", are generally spatially heterogeneous due to enhanced infiltration at the spatial locations where there is vegetation.  The unpredictable nature of rainfall is captured through  random timing and random overall strength of the kicks.

\subsection{The ``Flow"}
\label{section:flow}
This part of the model evolves the biomass density $B(X,T)$ $[kg/m^2]$ and the soil moisture $W(X,T)$ $[cm]$ for $T_i< T < T_{i+1}$  on the long timescale  between storms. It is initialized at $T_i$ with the $i$th post-storm soil water distribution ${\cal W}_i(X)$, which was enhanced by the kick $\Delta W_i(X)$ during the preceding storm.  
It's given by
\begin{subequations}
	\label{equationswitch2}
	\begin{align}
		\frac{\partial W}{\partial T}&=-LW-
		\Gamma B\Bigl(\frac{W}{1+W/A}\Bigr)\label{eqW2}\\
		\frac{\partial B}{\partial T} &= C
		\Gamma B\Bigl(\frac{W}{1+W/A}\Bigr)\Bigl(1-\frac{B}{K_B}\Bigr) -MB+D_B\frac{\partial^2 B}{\partial X^2}.
		\label{eqB2}
	\end{align} 
\end{subequations}
Here biomass dispersal is modeled as diffusion, and
we neglect soil water diffusion and leakage. The only water transport in the flow-kick model occurs via the kicks, described in Section~\ref{section:kick}. 
Soil moisture is lost through evaporation, at linear rate $L$,  and water uptake by vegetation, with characteristic rate $\Gamma$ and, following Reitkerk et al.~\cite{rietkerk2002self}, with a nonlinear dependence on soil water that limits the uptake rate as $W$ approaches a saturation level $A$.  (We note that this nonlinear saturation effect was neglected in the original pulsed precipitation model~\cite{gandhi2023pulsed}.) 
The uptake of soil water drives vegetation growth, up to its carrying capacity $K_B$, with water-use efficiency characterized by $C$. Plant mortality is taken into account through a linear biomass loss term with rate $M$.  The flow portion of the model does not incorporate positive feedbacks between biomass level and infiltration;  those are captured by the fast processes of the kick described next. Under the assumption that we are considering a small portion near the middle of a much larger hillslope, we apply periodic boundary conditions on a one-dimensional domain throughout.  
We discuss advantages and limitations of this choice in Section~\ref{sec:discussion}. 
Typical values for the parameters of the model, and the processes it captures,  are summarized in Table~\ref{tab:dim}. 

\subsection{The ``Kick"}
\label{section:kick}
We determine the kick $\Delta W(X)$ to the soil water by solving, via the method of characteristics, the system 
\begin{subequations}
	\label{equationswitch1}
	\begin{align}
		\frac{\partial H}{\partial T}&=
		-K_I\Bigl(\frac{{\cal B}(X)+fQ}{{\cal B}(X)+Q}\Bigr)\Theta(H)
		+
		V_0\frac{\partial}
		{\partial X}
		\Bigl(\frac{H}{1+N{\cal B}(X)}\Bigr)
		\label{eqH1} \\
		\frac{\partial W}{\partial T}&=
		K_I\Bigl(\frac{{\cal B}(X)+fQ}{{\cal B}(X)+Q}\Bigr)\Theta(H)\label{eqW1}
	\end{align} 
\end{subequations}
to its final steady state.
Here  $\Theta$ is the Heaviside step function and ${\cal B}(X)=B(X,T_i)$ is the biomass distribution at the time of the storm arrival.    The underlying assumption here is that the biomass does not evolve significantly on the timescale of the rain storm and can therefore be taken as fixed in time.  We also neglect evaporative losses during storms under similar reasoning that this loss will not be significant on this short timescale. 
We solve these equations with an initial uniform surface water height given by the storm depth $H_0$ $[cm]$ and determine the kick to soil moisture, $\Delta W(X)$, as the added water to the soil once all surface water has infiltrated, i.e. once the  Heaviside function reaches zero over the entire domain.

All of the positive feedbacks between biomass and soil moisture are captured by this kick part of the model.  With the default value of $f=0.1$ in~\eqref{equationswitch1}, the bare soil infiltration rate (${\cal B}=0$ $kg/m^2$) is a factor of 10 slower than its maximum rate, $K_I$.
This sigmoidal transition from low to high infiltration rate with increasing biomass ${\cal B}$ is an essential positive feedback, and suggests an advantage for the biomass level to exceed the threshold  $Q$ within patterned vegetation bands. 
The overland flow speed is also slower in regions where there is vegetation, which is another positive feedback that helps to concentrate the soil moisture where there is vegetation. We model the decrease in flow speed via the factor $(1+N{\cal B}(X))$ in the advection term in ~\eqref{eqH1}; if  ${\cal B}=Q$ then the speed is decreased by a factor of $3$ from the bare soil flow speed $V_0$ for the parameters of Table~\ref{tab:dim}. 
Throughout, we consider the uphill slope of the terrain to be uniform and in the $+X$-direction; the value of $V_0$ is estimated assuming a gradual $0.5\%$ grade, which is typical for the regions where vegetation bands occur~\cite{deblauwe2012determinants}.

\begin{table}[tbp]
\renewcommand{\arraystretch}{1.55}
\centering
\begin{tabular}{|c|c|c|l|} 
	\hline
	parameter & units& default value & description/definition \\
	\hline
	\hline
	${\cal H}_0$ & $cm$ & 1 & characteristic storm depth \\
	\hline
	$K_I$ & $ cm/day$ & 200 & infiltration rate coefficient  \\
	\hline
	$f$ & -- & 0.1 & bare/vegetated infiltration contrast\\
	\hline
	$Q$ & $kg/m^2$ & 0.1 & biomass level for infiltration enhancement \\
	\hline
	$V_0$& $m/day$ & $1.4\times 10^4$ & surface water speed (bare soil)\\
	\hline
	$N$ & $m^2/kg$& 20 & surface roughness coefficient\\
	\hline
	$L$ & $day^{-1}$ &0.0075& evaporation rate\\
	\hline
	$\Gamma$ & $(kg/m^2)^{-1}day^{-1}$ & 0.025 & transpiration coefficient\\
	\hline
	$A$ & $cm$ & 10  &  saturation level for transpiration \\
	\hline
	$C$ & $(kg/m^2)/cm$ &0.1 & water use efficiency coefficient\\
	\hline
	$K_B$ & $(kg/m^2)$ &4&  biomass carrying capacity
	\\
	\hline
	$M$& $day^{-1}$& 0.01 & biomass mortality rate\\
	\hline
	$D_B$& $m^2/day$& $0.01$ &  biomass diffusion coefficient\\
	
	\hline
	\hline
	$H_0$ & $cm$ & 1$^*$ 
	& mean storm depth\\
	\hline
	$T_d$ & $days$ & 15$^*$  
	& mean time between storms\\
	\hline
	\hline
	$\eta$& --& 2 & $\eta\equiv NQ$\\
	\hline
	$\alpha$ & --& 0.25 & $\alpha\equiv {\cal H}_0 C\Gamma/M$\\
	\hline
	$\sigma$ &--& 0.75& $\sigma\equiv L/M$\\
	\hline
	$\gamma$ &--&0.25 & $\gamma\equiv \Gamma Q/M$\\
	\hline
	$\kappa$&--& 40&$\kappa=K_B/Q$\\
	\hline
	$\zeta$& -- & 0.4& $\zeta\equiv M/(C\Gamma A)$\\
	\hline
	$\delta$ & -- &  0.0002 & $\delta\equiv D_BK_I^2/(M{\cal H}_0^2V_0^2)$
	\\
	\hline
\end{tabular}
\caption{\label{tab:dim}  Default parameter values for  the dimensioned kick-flow model~\eqref{equationswitch2}-\eqref{equationswitch1}, the random rainfall model described in Section~\ref{sec:model:rain} and the non-dimensionalized model \eqref{eq:slow:nondim}-\eqref{eq:fast:nondim}.  
	See~\cite{gandhi2023pulsed,gandhi2020fast} for further details on parameter value choices.\vspace{1mm}\\
	{\footnotesize $^*$While varying rainfall parameters $H_0$ or $T_d$, we typically use the default value listed here for the other one that remains fixed. }
} 
\end{table}%

\subsection{Rainfall Models}
\label{sec:model:rain}We assume that the Dirac-delta function storms instantaneously deposit water into the soil at some time $T_i$ with spatial distribution determined by the final steady-state solution to the system~\eqref{equationswitch1}.  We label the storm depth $H_0$ and the dry period between storms $T_d$.  We are interested in the impact of randomness, associated with kick timing and strength, on the resilience of these dryland ecosystems. In subsequent sections, we compare regions of existence and stability of different vegetation states -- spatially uniform and patterned -- between the idealized periodic rainfall case and the random rainfall case.  For periodic rainfall, we assume fixed values for $T_d$ and $H_0$, and the mean annual precipitation is given by $\MAP=H_0 T_{yr}/T_d$ where $T_{yr}=365\;days$. 
For random rainfall, we model the storm arrival times 
as a Poisson point process, and the total amount of water deposited by each storm is another random variable. Ours is a very simplified version of models for rainfall that have been developed in the hydrology literature; see, for example,~\cite{rodriguez1987some} and references therein. We consider the storm depths $H_0$, and duration of the dry periods between storms $T_d$ to each be exponentially distributed, with rate parameters $\lambda_H$ $[cm^{-1}]$,  $\lambda_T$ $[days^{-1}]$, 
\begin{eqnarray*}
H_0&\sim& Exp[\lambda_H],\qquad  \mathbbm E[H_0]=1/\lambda_H,\\
T_d&\sim& Exp[\lambda_T],\qquad \mathbbm E[T_d]=1/\lambda_T.
\end{eqnarray*}
The mean annual precipitation rate ($MAP$) is then the product of the mean number of storms in a year, $\mathbbm E[N_{storms}T_{yr}]=T_{yr}\lambda_T$, and the mean storm depth, $\mathbbm E[H_0]=1/\lambda_H$. Specifically, 
\begin{equation}
\label{eq:MAP1}
MAP
=T_{yr}\ \Bigl(\frac{\lambda_T}{\lambda_H}\Bigr)=T_{yr} \frac{\mathbbm E[H_0]}{\mathbbm E[T_d]}\ 
\end{equation} 
for the random rainfall model where, as before, $T_{yr}=365\;days$. 

\subsection{Dimensionless Parameters}
\label{sec:pulsedprecip:dim}
We non-dimensionalize in the same way as Gandhi et al.~\cite{gandhi2023pulsed}, with 
two different (dimensionless) timescales, $t$ and $\tau$ for the kick (fast) and flow (slow) subsystems, respectively, and a dimensionless distance $x$. Specifically, we let
\begin{equation}
\label{eq:nonddef1}
t=\frac{K_I}{{\cal H}_0}T,
\quad \tau=MT, \quad 
x=\frac{K_I/{\cal H}_0}{V_0}X.
\end{equation}
Here ${\cal H}_0$ is a characteristic storm depth and ${\cal H}_0/K_I$ is an associated infiltration timescale. This time, together with a characteristic overland flow speed $(V_0)$, determines a characteristic overland travel distance  $({\cal H}_0/K_I)/V_0$ that is used to non-dimensionalize $X$.  We set the (slow) biomass timescale by its mortality rate,  $M$.  Finally, we define the dimensionless fields:
\begin{equation}
\label{eq:nonddef2}
h=\frac{H}{{\cal H}_0},\quad w=\Bigl(\frac{C\Gamma}{M}\Bigr)W,
\quad b=\frac{B}{Q}.
\end{equation}

The non-dimensionalized flow subsystem of the model is
\begin{subequations}
\label{eq:slow:nondim}
\begin{align}
	\frac{\partial w}{\partial \tau}&= -\sigma w-\gamma b\Bigl(\frac{w}{1+\zeta w} \Bigr)\label{eq:slow:nondim:w}\\
	\frac{\partial b}{\partial \tau} &= b\Bigl(\frac{w}{1+\zeta w}\Bigr)\Bigl(1-\frac{b}{\kappa}\Bigr)-b+\delta \frac{\partial^2 b}{\partial x^2}.
	\label{eq:slow:nondim:b}
\end{align} 
\end{subequations}
The kick subsystem of the flow-kick model, in dimensionless variables, is
\begin{subequations}
\label{eq:fast:nondim}
\begin{align}
	\frac{\partial h}{\partial t}&= -\iota(x) \Theta(h) +\frac{\partial}{\partial x}\left( \nu(x) h
	\right)\label{eq:fast:nondim:h}\\
	\frac{\partial w}{\partial t} &= \alpha \iota(x) \Theta(h), \label{eq:fast:nondim:w}
\end{align} 
\end{subequations}
where
\begin{equation}
\label{eq:iotanu}
\iota(x)=\frac{\hat{b}(x)+f}{\hat{b}(x)+1},\quad  \nu(x)=\frac{1}{1+\eta \hat{b}(x)}.
\end{equation}
Because the rainstorm is assumed to deposit water on the surface instantaneously, we take $h=h_0$ as the initial condition for Equation~\eqref{eq:fast:nondim:h}. The dimensionless biomass distribution $\hat{b}(x)\equiv {\cal B}(X)/Q$ is taken from the flow system at the arrival time of the precipitation kick. 
Definitions of the dimensionless parameters, and typical values,  are given in Table~\ref{tab:dim}.

\subsection{Numerical methods}
As detailed in Section~\ref{supp:slovefast} of the Supplement, which is based on the approach developed in~\cite{gandhi2023pulsed}, we can use the method of characteristics to obtain the following integral expression for the soil water kick $\Delta w_i(x)$ from the $i$th rain storm occurring at $\tau_i$ with depth $h_i$: 
\begin{equation}
\label{eq:nm:Omegax0}
\Delta w_i(x)=\alpha\iota(x)
\int_{x}^{y_\ell(x)}\frac{\Theta\Bigl(h_i-\hat{h}(y;y_\ell(x))\Bigr)}{\nu(y)} dy,
\end{equation}
where 
\begin{equation}
\label{eq:nm:hhat0}
\hat{h}(y;y_\ell(x))=\frac{1}{\nu(y)}\int_x^y\iota(s) ds.    
\end{equation}
The infiltration rate $\iota(x)$ and flow speed $\nu(x)$ functions have spatial dependence prescribed by the current biomass distribution $\hat{b}(x)$ via \eqref{eq:iotanu}.
The integral~\eqref{eq:nm:hhat0}  represents the height of the surface water at a point $y$ along a characteristic that starts at $x$ with $h=0$ and follows it backward in time to the largest $y$ where $\hat{h}=h_i$, denoted  $y_\ell(x)$. 

To numerically simulate the flow-kick model, we discretize the slow flow~\eqref{eq:slow:nondim} in space using centered finite-differences and employ a fourth-order, variable timestep Runge-Kutta solver from Matlab's ODE suite~\cite{shampine1997matlab} for integration over time intervals $(\tau_{i-1},\tau_i)$.  At a time $\tau_i$, with storm of depth $h_i$ arriving, we numerically integrate equations~\eqref{eq:nm:Omegax0}-\eqref{eq:nm:hhat0} via the trapezoid rule to compute the soil water kick $\Delta w_i(x)$ from the given storm. We use linear interpolation to determine the boundary of the integral in equation~\eqref{eq:nm:Omegax0} imposed by the Heaviside step function~$\Theta$ where $\hat{h}=h_i$. We then integrate the slow flow model~\eqref{eq:slow:nondim} with the pre-storm biomass $\hat{b}$ and updated soil water $w\to w+\Delta w_i$ as initial conditions for the next iteration over the time interval $(\tau_i,\tau_{i+1})$, where $\tau_{i+1}$ is the arrival time for the next storm.  Under the assumption that we are modeling a small section in the middle of a long hillslope covered by vegetation patterns, we incorporate periodic boundary conditions throughout.   Unless otherwise stated we use a 1~$km$ domain with a spatial grid spacing of $\Delta x=0.2\, m$.  While we perform computations with the dimensionless model, we will typically report results using the associated dimensioned quantities. 

\begin{figure}
\label{fig:SimExample}
\includegraphics[width=\textwidth]{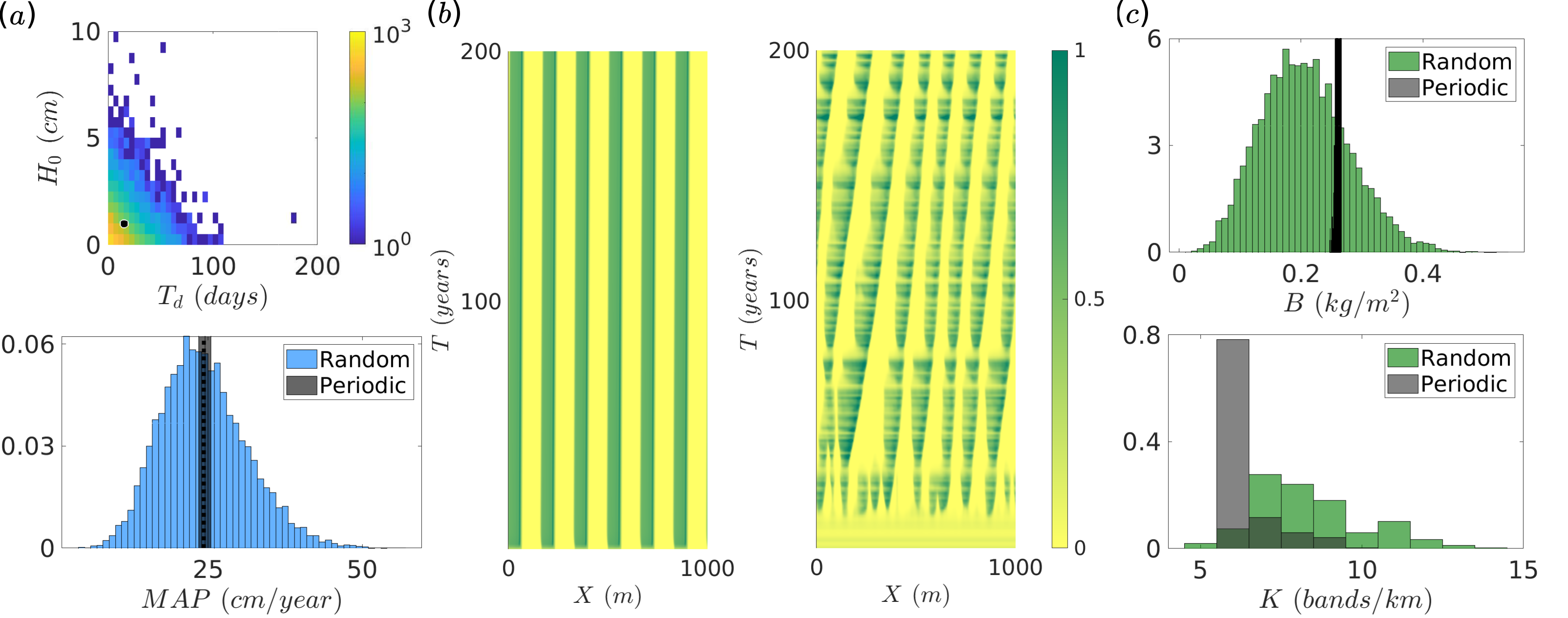}
\centering
\caption{(a) Distribution of storms as a function of mean depth $H_0$ and mean duration $T_d$, and the distribution of mean annual precipitation $MAP$.  (b) Spacetime plots of biomass from simulations initialized with the same noise on top of a uniform state and run  with periodic and random rainfall.   (c)  Distribution of domain-averaged biomass $B$ and  wavenumber $K$ from the last 100 years of 100 trials with the same parameters.  
	\label{fig:nm:example}
}
\end{figure}

In Figure~\ref{fig:nm:example}, we present results from simulations of the flow-kick model described by~\eqref{eq:slow:nondim} and \eqref{eq:nm:Omegax0} with standard parameters listed in Table~\ref{tab:dim}.   We compare periodic rainfall with storms of depth $H_0=1\,cm$ and dry periods of $T_d=15\,days$ to random rainfall with the same mean storm characteristics.  We take 100 trails initialed with 1\% uniform random noise on top of a spatially uniform state and run the simulation with both rainfall models for 200 years. Figure~\ref{fig:nm:example}(a) shows rainfall statistics and Figure~\ref{fig:nm:example}(b) shows spacetime plots of the annually-averaged biomass from one of the trials for the periodic rainfall case (left) and for the random rainfall case (right).  In Figure~\ref{fig:nm:example}(c), histograms of the domain-average biomass $B$ and wavenumber $K$ are shown for the final century over all 100 trials.   The simulations with periodic rainfall produce traveling wave solutions with wavenumbers ranging from 6 to 10 $bands/km$, but the distribution is strongly peaked at $K= 6\, bands/km$.  With random rainfall, the mean wavenumbers for the pattern has a peak of 7 $bands/km$ and is more broadly distributed.

For the examples shown in Figure~\ref{fig:nm:example}(b), the traveling wave pattern with periodic rainfall is nearly stationary, while the bands in the pattern with random rainfall clearly exhibit uphill migration. 
We find that the spatial resonance phenomenon~\cite{gandhi2023pulsed} underlying pattern formation in the idealized periodic rainfall case leads to a migration speed that is highly sensitive to the pattern wavelength and the rainfall parameters. For example, as seen in Section~\ref{sec:sim} and discussed further in~\cite{gandhi2023pulsed}, some patterns obtained with nearby rainfall parameters, or with slightly different wavenumbers, may even switch between migrating uphill and migrating downhill. 

\section{Spatially uniform states and the transcritical bifurcation point}
\label{sec:transcrit}

In this section, we focus on conditions for existence of a spatially uniform, 
vegetated state which we refer to as the \textit{uniform vegetation state}. We are specifically interested in determining the threshold for this state to exist in the flow-time ($T_d$), kick-size ($H_0$) parameter plane. Below this threshold, we find that the only spatially-homogeneous state is the zero-biomass one, which we refer to as the \textit{bare soil state}. We find that the threshold corresponds to a transcritical bifurcation point wherein the bare soil state loses stability as the uniform vegetation state emerges. We investigate the impact of randomness on this threshold, in the case where the dimensionless storm depths $h_0$ and/or dry periods $\tau_d$ are drawn from exponential distributions.   
We demonstrate that this stochasticity shifts the boundary so that larger, on average, and/or more frequent, on average, kicks are required to support uniform vegetation.

\begin{figure}
\includegraphics[width=15cm]{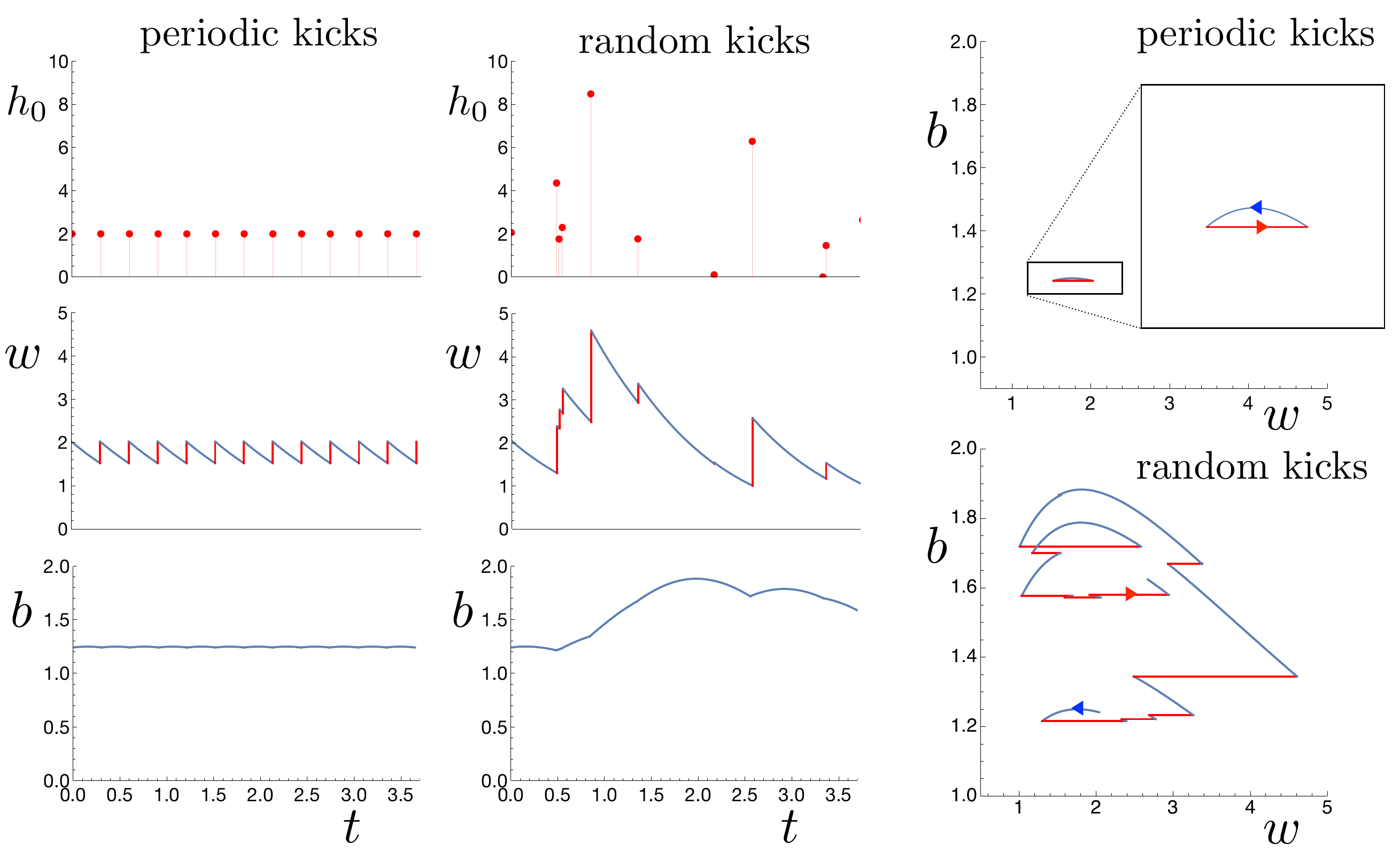}
\centering
\caption{Two examples of a sequence of surface water kicks $h_0$, periodic vs. random, along with the associated time series for soil moisture $w(t)$, biomass $b(t)$, and associated phase-space representations, obtained from~\eqref{eq:flowuniform}. The kicks $\Delta w$ are indicated as red segments, and flows in blue. A blow-up of the small closed orbit in the phase plane in the periodic case is shown, upper-right; the trajectory in the random kick case wanders over a relatively large region of the phase space. In the periodic case, there are 12 equally-spaced, identical storm kicks in a year. For the random case, the kicks and their spacing are drawn from exponential distributions with the same mean values as the periodic case. All other parameters are given in Table~\ref{tab:dim}.
	\label{fig:timeseries}
}
\end{figure}

The flow, defined by  
\begin{equation}
\label{eq:phiflowdef}
\phi_{\tau}(w_0,b_0)=(w(\tau),b(\tau)),    
\end{equation} 
is determined by solving \eqref{eq:slow:nondim}, from an initial condition $(w_0,b_0)$, out to  time $\tau$. In the spatially uniform case, this means solving the ordinary differential equations
\begin{eqnarray}
\label{eq:flowuniform}
\dot{w}&=& -\sigma w-\gamma b\Bigl(\frac{w}{1+\zeta w} \Bigr)\nonumber\\
\dot{b} &=& b\Bigl(\frac{w}{1+\zeta w}\Bigr)\Bigl(1-\frac{b}{\kappa}\Bigr)-b.
\end{eqnarray}
The kick $\Delta w_0$ to the soil moisture, that results from an initial surface water height $h_0$, is determined from
\eqref{eq:fast:nondim} to be simply $\Delta w_0=\alpha h_0$
in the spatially uniform case. Thus a complete flow-kick cycle, takes an initial condition $(w_0,b_0)$ to the state $(w_1,b_1)$, where
\begin{equation}
\label{eq:uniformmap}
(w_1,b_1)=\phi_{\tau_d}(w_0,b_0)+(\Delta w_0,0).
\end{equation}
Examples are shown in Figure~\ref{fig:timeseries} for both the case of periodic kicks and random kicks.

\begin{figure}
\includegraphics[width=15.5 cm]{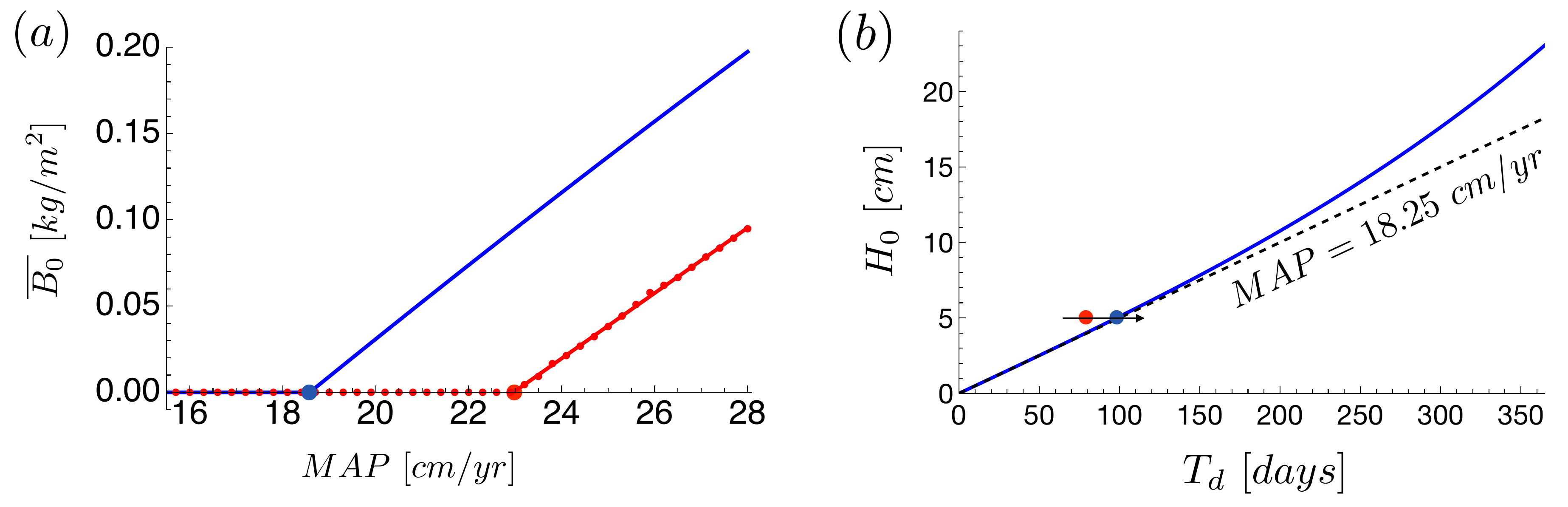}
\centering
\caption{(a) Plot of biomass level $B_0$ at the start of a flow-kick cycle vs. mean annual precipitation rate ($MAP$). The blue solid curve is for the case of periodic kicks with $H_0=5\ cm$, and incrementing the time between kicks to decrease $MAP$. There is a transcritical bifurcation point indicated by the blue dot at $MAP\approx 18.6 \ cm/year$. In the random kick case, the (red) dots were obtained by averaging $B_0$ over $5\times 10^5$ cycles of the flow-kick for each data point. The red line is a linear fit to the last 16 data points. We find that the biomass level drops to zero around $23\ cm/year$ in the random case. (b) A plot of the transcritical bifurcation point \eqref{eq:detthreshold} in the periodic case, in the ($T_d,\ H_0)$ parameter plane. The blue and red dots are for the values from (a), associated with $H_0=5\ cm$, and the arrow through those points indicates the parameter path taken, with increasingly long dry periods, leading to a decrease in $MAP$ from 28 $cm/year$ to 16 $cm/year$.
	\label{fig:transcritical}
}
\end{figure}

The mean annual precipitation rate is $MAP=3.65\ h_0/\tau_d$  $[cm/year]$ in the periodic rainfall case; in the random rainfall case, $h_0$ and $\tau_d$ are replaced by their mean values as in \eqref{eq:MAP1}. If $MAP$ is too low then the biomass collapses. However, how low it can go before collapse, within the flow-kick model, depends on the details of the rainfall model. An indication of this is provided by
Figure~\ref{fig:transcritical}. It shows a bifurcation diagram of biomass level, at the start of a flow-kick cycle ($B_0=Qb_0$), as a function of the mean annual precipitation rate for the case of periodically applied $H_0=5 cm$ kicks (blue bifurcation curve). This is compared to a case of random kicks (red points), where the mean biomass level $\overline{B_0}$ is obtained by averaging over $5\times 10^5$ cycles of the flow-kick system. We see decreasing mean biomass, with decreasing $MAP$, that eventually reaches a zero biomass level at a point of stabilization of the unproductive state, as happens in a transcritical bifurcation. In this example, we also see that the bifurcation point has shifted to a higher $MAP$ in the random rainfall case, compared to the periodic rainfall model. 

In order to determine the threshold rainfall level needed to sustain the ecosystem, as a function of rainfall parameters, we focus on the flow equations \eqref{eq:flowuniform}, linearized about $b=0$. We are interested in determining whether a small perturbation $\delta b_0$ grows or decays over the course of flow-kick cycles. Specifically, we estimate $\phi_{\tau_d}(w_0,\delta b_0)$ from \eqref{eq:flowuniform}, linearized about $b=0$, i.e. we solve
\begin{equation*}
{\dot\delta b}= \Bigl(\frac{w(t)}{1+\zeta w(t)}-1\Bigr)\ \delta b,
\end{equation*}
where the base state for the perturbation is \begin{equation*}
w(t)=w_0 e^{-\sigma t}.
\end{equation*}
This linearized problem yields
$\phi_{\tau_d}(w_0,\delta b_0)= (\mu w_0,\delta b_1)$,
where
\begin{equation}
\label{eq:growth}
\ln \Bigl(\frac{\delta b_1}{\delta b_0}\Bigr)=\Bigl(\frac{1}{\sigma\zeta}\Bigr)\ln \Bigl(\frac{1+\zeta w_0}{1+\zeta \mu w_0}\Bigr)-\tau_d;\quad   \mu \equiv e^{-\sigma \tau_d}\in (0,1).
\end{equation}
One cycle of flow-kick leads to the new uniform state, approximated by $(w_1,\delta b_1),$ where $w_1=\mu w_0+\Delta w_0$, and $\delta b_1$ is determined from \eqref{eq:growth}. Whether the perturbation $\delta b_0$ grows or decays in one cycle is determined by the sign of the right-hand-side of \eqref{eq:growth}. 

We first determine the threshold in the deterministic case, where every kick is identically $\Delta w_0=\alpha h_0$, and precisely timed at intervals  $\tau_d$. Here the biomass perturbation $\delta b_0$ is applied to the periodic, zero biomass solution, which satisfies $w_1=w_0$ in \eqref{eq:uniformmap}, where $w_1=\mu w_0+\Delta w$. Substituting $w_0=\Delta w_0/(1-\mu)$ into \eqref{eq:growth}, we determine the threshold for biomass perturbations to grow by setting the right-hand-side to zero. This yields
\begin{equation}
\label{eq:detthreshold}
\Delta w_0^{det}=\frac{(1-\mu^\zeta)(1-\mu)}{\zeta(\mu^\zeta-\mu)},
\end{equation}
where  $\zeta<1$ is required for biomass to possibly grow. This bifurcation set is shown  in Figure \ref{fig:transcritical}(b), for the default parameters of Table~\ref{tab:dim}.  Below it, uniform vegetation collapses to the zero biomass state; above it, vegetation is sustainable. 

We next examine how the threshold shifts, as suggested by Figure~\ref{fig:transcritical}, when the kicks and/or their timing are random variables. 
We proceed by estimating the expected value of the growth rate of biomass perturbations $\delta b_0$, given by \eqref{eq:growth}:
\begin{equation}
\mathbbm E\Bigl[\ln\Bigl(\frac{\delta b_1}{\delta b_0}\Bigr)\Bigr]=\Bigl(\frac{1}{\sigma\zeta}\Bigr)\mathbbm E\Bigl[\ln\Bigl(\frac{1+\zeta w_0}{1+\zeta\mu w_0} \Bigr)\Bigr]-\mathbbm E[\tau_d].
\label{eq:ferandom}
\end{equation}
Vegetation perturbations to the zero-biomass state  grow, on average, if 
\begin{equation}
\mathbbm E\Bigl[\ln\Bigl(\frac{1+\zeta w_0}{1+\zeta\mu w_0} \Bigr)\Bigr]>\sigma\zeta \mathbbm E[\tau_d],
\label{eq:ferandom-2}
\end{equation}
where $w_0$ is a random variable, as is $\mu=e^{-\sigma \tau_d}$ when the timing of kicks are random.

\begin{figure}

\includegraphics[width=15cm]{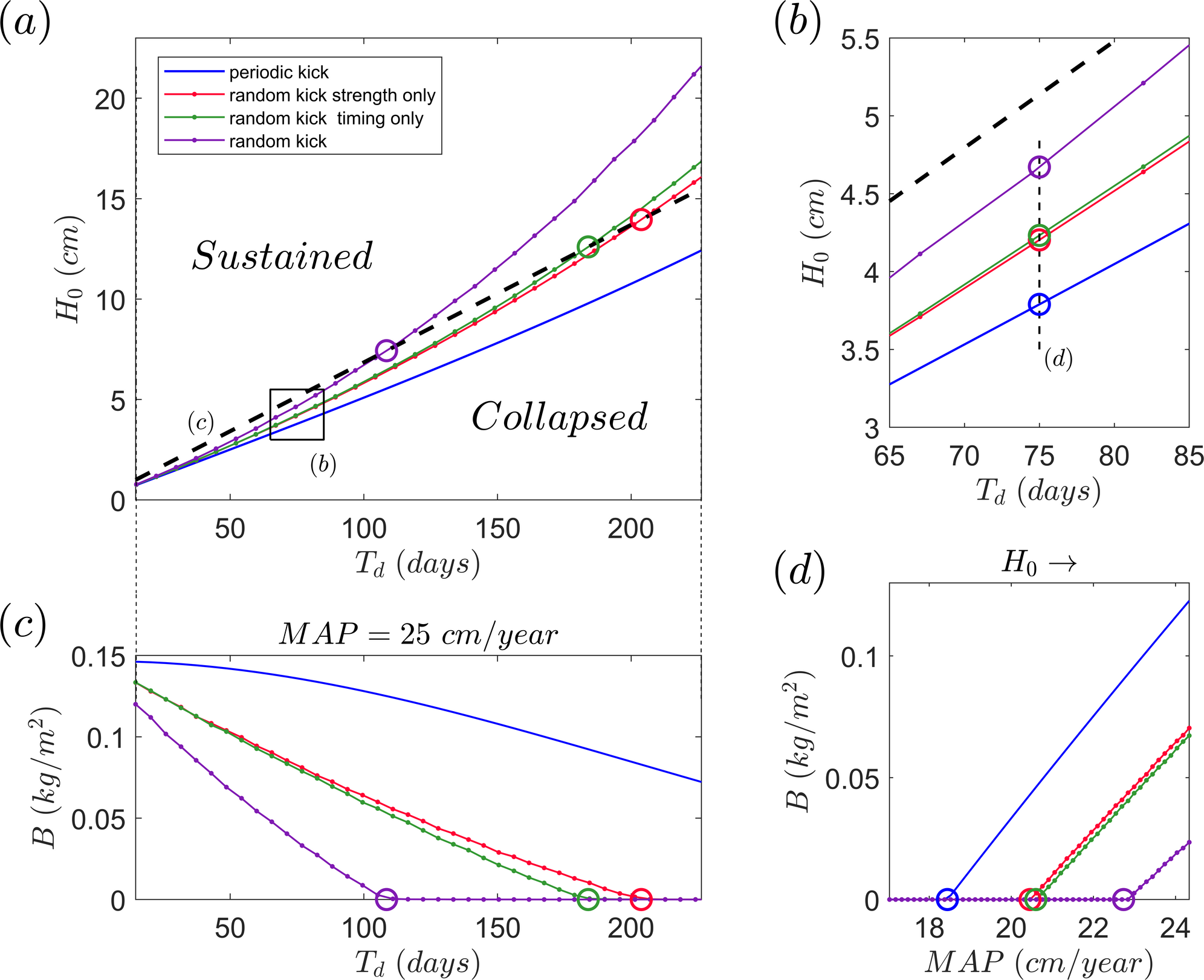}
\centering
\caption{(a) Zero-biomass stability boundaries for periodic kick, random kick strength only, random kick timing only, and fully random kick cases. Below the curves the zero-biomass (collapsed) state is stable with respect to uniform perturbations. (b) Zoom-in of black rectangle in panel (a). (c) Average biomass vs. $T_d$ with mean annual precipitation fixed at 25 cm/year along dashed line in panel (a). (d) Average biomass vs mean annual precipitation with $T_d$ fixed at 75 days along dashed line in panel (b). In panels (c) and (d), the circles indicate the bifurcation points predicted by the intersection of the dashed lines in (a) and (b) with the stability boundaries.
	\label{fig:uniformthreshold}
}
\end{figure}

Figure~\ref{fig:uniformthreshold} summarizes our results for the threshold in the $(\mathbbm E[T_d],\mathbbm E[H_0])$-parameter plane for sustaining a uniform vegetation state. Above the thresholds in panel (a), the zero-biomass state is unstable with respect to uniform perturbations, so we obtain a sustained uniform vegetation state. 
When the kicks are periodic with identical strength, the threshold is given explicitly by  \eqref{eq:detthreshold}. We also consider three random cases: only the kick strength is a random variable ($H_0\sim Exp[\lambda_H]$), only the kick  timing  is a random variable  ($T_d\sim Exp[\lambda_T]$), and finally both kick strength and  timing are random variables ($H_0\sim Exp[\lambda_H]$ and $T_d\sim Exp[\lambda_T]$). For the three random cases, the thresholds correspond to an equality in  \eqref{eq:ferandom-2}, where the expectation on the left-hand-side is over the distributions of post-kick water $w_0$ and $\mu=e^{-\sigma\tau_d}$. In the case where only the random kick strength is a random variable, we are able to find an analytic expression for this expectation (see the Supplementary Materials for more details). By comparing this expectation to a brute force estimation of the left-hand side of \eqref{eq:ferandom-2} over many kick-flow cycles, we find $10^6$ cycles to be sufficient to compute the expectation to 3 digits of accuracy. We use this number of cycles to compute the expectation numerically for the other two random cases, and find where it equals the right-hand-side of \eqref{eq:ferandom-2} via a bisection search. Panels (c) and (d) provide numerical validation  of the bifurcation points  predicted by these thresholds: by tracing the two dashed lines in the $(\mathbbm E[T_d],\mathbbm E[H_0])$-parameter plane and recording the mean biomass over $10^6$ flow-kick cycles, we see that the intersections with the thresholds (indicated by the circles) accurately predict the transcritical bifurcation points where the uniform vegetation state reaches zero. 

Interestingly, panel (c) shows that for a fixed mean annual precipitation, how that rain is distributed in time can affect whether or not the uniform vegetation state collapses or is sustained. For example, when the mean annual precipitation is $\MAP=25\,cm/year$ with $T_d=150\, days$, the uniform vegetation state has collapsed in the fully random case, whereas the uniform state is sustained in the other three cases. We also notice that the boundaries for the random kick strength only and random kick timing only are very similar, while when both kick strength and kick timing are random, the offset from the deterministic case is approximately doubled compared to when just one is random.

\section{Pattern Formation}
\label{sec:pattern}
In this section we explore the impact of rainfall variability on pattern formation in the flow-kick model of vegetation described in Section~\ref{sec:kickflow}.  We characterize linear stability of the uniform state that emerges from the transcritical point observed in Section~\ref{sec:transcrit} under both periodic and random rainfall.  We construct a descrete-time map that describes the dynamics of small spatially-periodic perturbations to the uniform states over a kick-flow cycle in Section~\ref{sec:pattern:Mmap}. We characterize stability of uniform states in terms of Lyaponuv exponents of this map in Section~\ref{sec:pattern:instability}.  Finally, in Section~\ref{sec:pattern:threshold}, we summarize our linear stability results in the flow-kick parameter plane.     We find that randomness in the timing and strength of the kicks~\eqref{eq:nm:Omegax0} to the soil water significantly decreases the parameter range over which the uniform state is unstable to spatially periodic perturbations.   In Section~\ref{sec:sim}, we validate these linear predictions by performing numerical simulations that  let  us study the resulting fully nonlinear patterns.

\subsection{Kick-Flow Stability Map} 
\label{sec:pattern:Mmap}

Here we describe the construction of a map which tracks the dynamics of perturbations of spatially uniform solutions of the pulsed-precipitation model over a kick-flow cycle.  We characterize stability by whether or not the amplitudes of these perturbations approach zero as the number of cycles tends towards infinity, which we determine by numerical approximation of the associated maximal Lyaponuv exponent.  With periodic rainfall, this provides identical  results to the linear stability analysis based on Floquet theory that is used in~\cite{gandhi2023pulsed}.

We assume some initial soil water and biomass $(w_0,b_0)$ and construct the spatially uniform kick-flow map
defined by
\[
\psi_{\tau_d,h_0}(w_0,b_0)\equiv \phi_{\tau_d}(w_0,b_0) + (\alpha h_0, 0),
\]
for rainstorm kick with storm depth $h_0$ and flow $\phi_{\tau}$ defined by equation~\eqref{eq:phiflowdef}.  We are interested in the dynamics of small perturbations of the associated spatially uniform trajectory of the model and thus consider a state of the form 
\begin{equation*}
(w_k(t),b_k(t))=\phi_t(w_0,b_0)+ (\delta w_k(t), \delta b_k(t))e^{ikx},
\end{equation*} 
where $k$ represents the dimensionless wavenumber of the perturbation.  We approximate the dynamics of the $k$-dependent perturbation amplitudes $\delta w_k(t)$ and $\delta b_k(t)$ from the linearized pulsed-precipitation model with flow~\eqref{eq:slow:nondim} and kicks~\eqref{eq:nm:Omegax0}.
Specifically, after one kick-flow cycle, the perturbation amplitudes  are given by
\begin{equation}
\label{eq:FlowKickPertMap}
\begin{pmatrix}
	\delta w_{k}(\tau_d)\\
	\delta b_{k}(\tau_d)
\end{pmatrix}=\mathcal{M}_{k}[w_0,b_0,h_0,\tau_d]
\begin{pmatrix}
	\delta w_{k}(0)\\
	\delta b_{k}(0)
\end{pmatrix}.
\end{equation}
where the kick-flow stability map,
\begin{equation*}
\mathcal{M}_k[w_0,b_0,h_0,\tau_d]=\Psi_k[w_0,b_0,h_0,\tau_d]\ \Omega_k[b_0,h_0],
\end{equation*} 
can be decomposed into a contribution $\Omega_k[b_0,h_0]$ from the rainstorm kick
of storm depth $h_0$, 
and a contribution $\Psi_k[w_0,b_0,h_0,\tau_d]$  from the slow flow during the following dry period of duration $\tau_d$.

The rainstorm affects only the soil water within the pulsed precipitation model, and can thus only contribute to changes in the $\delta w_k$ perturbation amplitude.  Moreover, soil water perturbations 
have no influence on spatial distribution of the soil water kick,  whereas biomass perturbations do affect this through positive feedbacks on flow speed and infiltration rate. 
The matrix associated with the kick then takes the form
\[
\Omega_k[b_0,h_0]=\begin{pmatrix}
1 & J_k[b_0,h_0]
\\    
0&1 
\end{pmatrix}.
\]
The full details for computing $J_k[b_0,h_0]$, given by  
\begin{equation}
\label{eq:Jk}
J_k[b_0,h_0] =\alpha h_0 \left[ \frac{1}{\iota(b)}\frac{d \iota}{d b}
+ \frac{1}{\nu(b)}\frac{d \nu}{d b} e^{ik\ell_0}
+\frac{i}{k \ell_0}\left(\frac{1}{\iota(b)}\frac{d \iota}{d b} + \frac{1}{\nu(b)}\frac{d \nu}{d b}\right)\left(e^{ik\ell_0}-1\right)
\right]\bigg\vert_{b=b_0},
\end{equation}
where
\begin{equation*}
\ell_0=\frac{\nu(b)}{\iota(b)}h_0, \quad \frac{1}{\iota(b)}\frac{d \iota}{d b}=\frac{(1-f)}{(b+f)(b+1)},\quad \text{and}\quad      \frac{1}{\nu(b)}\frac{d \nu}{d b}=-\Bigl(\frac{\eta}{1+\eta b}\Bigr),
\end{equation*}
are provided in~\cite{gandhi2023pulsed}.

The matrix $\Psi_k[w_0,b_0,h_0,\tau_d]$, associated with the wavenumber $k$ perturbation, is computed by integration of the slow flow~\eqref{eq:slow:nondim} linearized about the spatially uniform, post-kick state $\phi_{\tau_d}(w_0+\alpha h_0,b_0)$. It is a solution of the matrix equation
\[\dot{\Psi}=\mathcal{A}_k(\tau) \Psi\]
to time $\tau=\tau_d$, with the identity matrix as the initial condition, and  where
\[
\mathcal{A}_k(\tau)=\begin{pmatrix}
-\sigma - \frac{\gamma b}{(1+\zeta w)^2} & 
-\frac{\gamma w}{ 1+\zeta w}\\
\frac{b}{ (1+\zeta w)^2} (1-\frac{b}{\kappa}) & 
-(\delta  k^2 + 1) + \frac{ w}{ 1+\zeta w}\left(1-2\frac{b}{\kappa}\right)
\end{pmatrix}\Bigg\vert_{(w,b)=\phi_\tau(w_0+\alpha h_0,b_0)}.
\] 
\subsection{Pattern Forming Instability}
\label{sec:pattern:instability}

We now use the flow-kick map defined by~\eqref{eq:FlowKickPertMap} 
to explore linear predictions about the pattern-forming instability of the spatially uniform vegetation state under both periodic and random rainfall.  With periodic rainfall, as discussed in~\cite{gandhi2023pulsed}, the wavelength of the unstable modes are tied to spatial resonances with a distance $\ell_0$, which is defined below~\eqref{eq:Jk}. Specifically, $\ell_0$ characterizes the distance storm water, of depth $h_0$, travels on the surface before infiltrating into the soil.  

We characterize stability of the uniform vegetation state of the pulsed precipitation model using the maximal Lyapunov exponent associated with the discrete dynamical system defined by~\eqref{eq:FlowKickPertMap}, that is:
\begin{align}
\label{eq:LyapExp}
\lambda_k &= \lim_{\tau\to\infty} \frac{1}{\tau}\ln\frac{\left\Vert\left(\delta w_k(\tau),\delta b_k(\tau)\right)\right\Vert}{\left\Vert\left(\delta w_k(0),\delta b_k(0)\right)\right\Vert}
\\[10pt] 
& =\lim_{n\to\infty}\left(\sum_{j=1}^n\tau_d^{(j)}\right)^{-1}\ln\left\Vert\prod_{j=1}^n\mathcal{M}_k\left[w_{j-1},b_{j-1},h_0^{(j)},\tau_d^{(j)}\right]
\right\Vert    \nonumber
\end{align}
where $\displaystyle \left(w_{j},b_{j}\right)=\phi_{\tau_d^{(j)}}\left(w_{j-1}+\alpha h_0^{(j)},b_{j-1}\right)$. The uniform state flow of the slow system $\phi_\tau$ is defined by equation~\eqref{eq:phiflowdef}
and we use the $L_2$ matrix norm for the ordered matrix product in the lower expression of~\eqref{eq:LyapExp}. 

\begin{figure}[h]
\centering
\includegraphics[width=\textwidth]{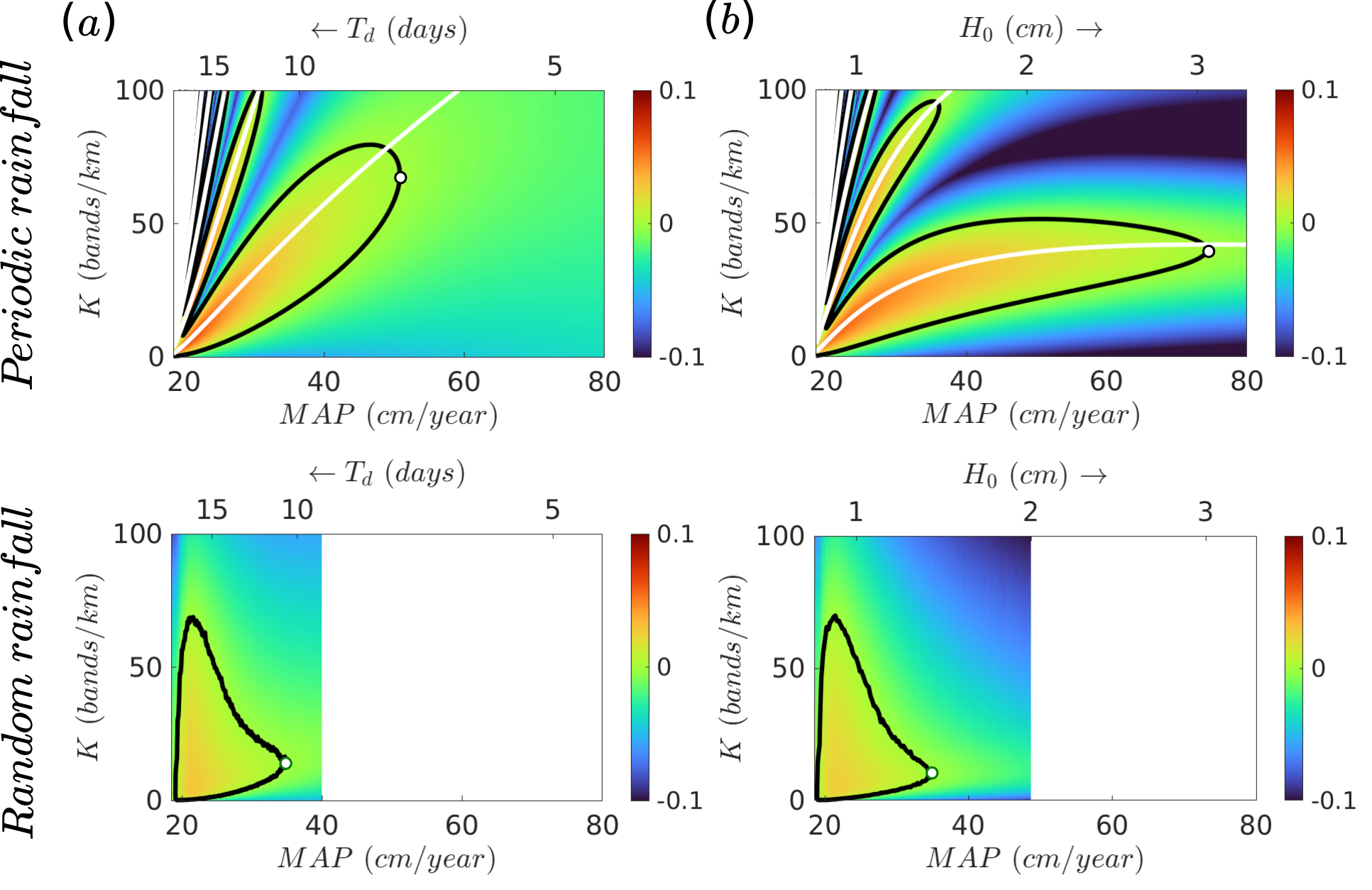}\caption{Maximal Lyaponuv exponent $\lambda_k$ given by equation~\eqref{eq:LyapExp} as a function of \MAP\ and (dimensioned) $K$. We control \MAP\ by (a) fixing mean storm depth at $H_0=1\, cm$ and varying the mean time between storms $T_d$, and by (b) fixing the mean time between storms to be $T_d=15\, days$ and varying mean storm depth $H_0$.   Solid black lines are linear stability boundaries with periodic rainfall (top) random rainfall (bottom).  White lines in the top row indicate predicted resonances $k_n^*$ between the pattern wavelength and characteristic surface flow distance with periodic rainfall. White circles indicate onset of instability and the corresponding \MAP\ values are indicated  by a transition in stability of the uniform vegetation state in Figure~\ref{fig:sim:intro}.  
}
\label{fig:patterns:mapk}
\end{figure}

\paragraph{Periodic Rainfall} In the case of periodic rainfall, we have fixed, identical values of $h_0$ and $\tau_d$ for each cycle, and the eigenvalues of the matrix $\mathcal{M}_k[w_0,b_0,h_0,\tau_d]$ are exactly the Floquet multipliers computed in~\cite{gandhi2023pulsed} to determine linear stability. The top row of Figure~\ref{fig:patterns:mapk} shows a heat map of $\lambda_k$, as a function of \MAP\ and the dimensioned wavenumber $K$ of the perturbation, for periodic rainfall varied in two distinct ways: (a) by fixing storm depth to $H_0=1\, cm$ and varying the interstorm time interval $T_d$, and (b) by fixing $T_d=15\, days$ and varying $H_0$.  The $\lambda_k=0$ instability boundary is indicated by a solid black curve, and in both cases we see the resonance  structure observed in~\cite{gandhi2023pulsed}. The center line of each resonance tongue is well approximated by the white lines defined, in dimensionless units, by $k^*_n=n\pi/\ell_0$ for $n=1,3,5\dots$ 
For the perturbation with wavenumber $k^*_n$, the distance between a maximum (corresponding to a newly-forming vegetation band) and a minimum (corresponding to a newly-forming bare soil region) is  exactly $\ell_0/n$. 
This agreement suggests that the preferred pattern wavelength, at least at the linear level, is the one where water from newly-forming bare soil regions is harvested by newly-forming vegetation bands.

With varying $T_d$ at fixed $H_0=1\,cm$, as shown in the top panel  of Figure~\ref{fig:patterns:mapk}(a), the uniform vegetation state remains stable for $\MAP> 52.4\,cm/year$, where lowest-order ($n=1$) resonance tongue first loses stability to perturbations with dimensioned wavenumber $K=68\, bands/km$.       
We see the resonance tongues pushed to smaller wavenumbers at a given \MAP\ in the top panel of Figure~\ref{fig:patterns:mapk}(b) where the rainfall is still periodic, but \MAP\ is varied by increasing $H_0$ instead of decreasing $T_d$. (This is consistent with $L_0$ setting preferred wavelengths since larger storm depths mean further distance traveled by surface water before infiltrating during storms.)  The onset of the pattern forming instability in this case is at $MAP=75.4$ with a wavenumber of $K=40\, bands/km$.  We note that the fastest growing modes predicted by this linear analysis with periodic rainfall have very short wavelengths relative to typical banded vegetation patterns as shown, for example, in Figure~\ref{fig:intro}.

\paragraph{Random Rainfall}  With random rainfall the largest Lyaponuv exponent $\lambda_k$, given by~\eqref{eq:LyapExp}, can be numerically approximated by computing perturbation amplitudes after $n$ repeated iterations of the flow-kick stability map~\eqref{eq:FlowKickPertMap} with kicks $h_0^{(j)}$ and dry periods $\tau_d^{(j)}$ drawn from exponential distributions with means $h_0$ and $\tau_d$, respectively. The approximation is given by  
\begin{equation}
\label{eq:LyapExpNum}
\lambda_k\approx \lambda_k^{(n)}\equiv \left(\sum_{j=1}^n\tau_d^{(j)}\right)^{-1}\ln\frac{\left\Vert\left(\delta w_k^{(n)},\delta b_k^{(n)}\right)\right\Vert}{\left\Vert\left(\delta w_k^{(0)},\delta b_k^{(0)}\right)\right\Vert}. 
\end{equation}
In the calculations described here, we typically take $n=10^{5}$. 
In order to reduce numerical rounding error with large $n$, we rescale by the norm of the perturbation amplitude after $m=10^{2}$ iterations as described by, for example, Meiss~\cite{meiss2007differential}.

The bottom row of Figure~\ref{fig:patterns:mapk} shows the maximal Lyapunov exponent for random rainfall as a function of the perturbation wavenumber $K$ and \MAP\ with (a) fixed mean storm depth $H_0=1\,cm$ and (b) fixed mean dry period between storms $T_d=15\,days$.  
Overall, the linear predictions based on the maximal Lyaponuv exponent of the flow-kick stability map~\eqref{eq:FlowKickPertMap}  show that randomness in rainfall storm depths and arrival times  impacts pattern forming instability in two significant ways: (1) the resonance structure observed with periodic rainfall disappears when randomness is introduced, and (2) the instability bubble in the $(\MAP,K)$ space shown in Figure~\ref{fig:patterns:mapk} shrinks significantly in size. This suggests that patterns, which are often thought of as a way to increase resilience, appear at much lower precipitation levels and in smaller diversity of wavenumbers as a result of rainfall variability.  In order to test these predictions and explore how they relate to the fully nonlinear patterns predicted by the model, we turn to numerical simulation in Section~\ref{sec:sim}.

\subsection{Stability Boundaries in Flow-Kick Parameter Space}
\label{sec:pattern:threshold}
\begin{figure}[h]
\centering
\includegraphics[width=\textwidth]{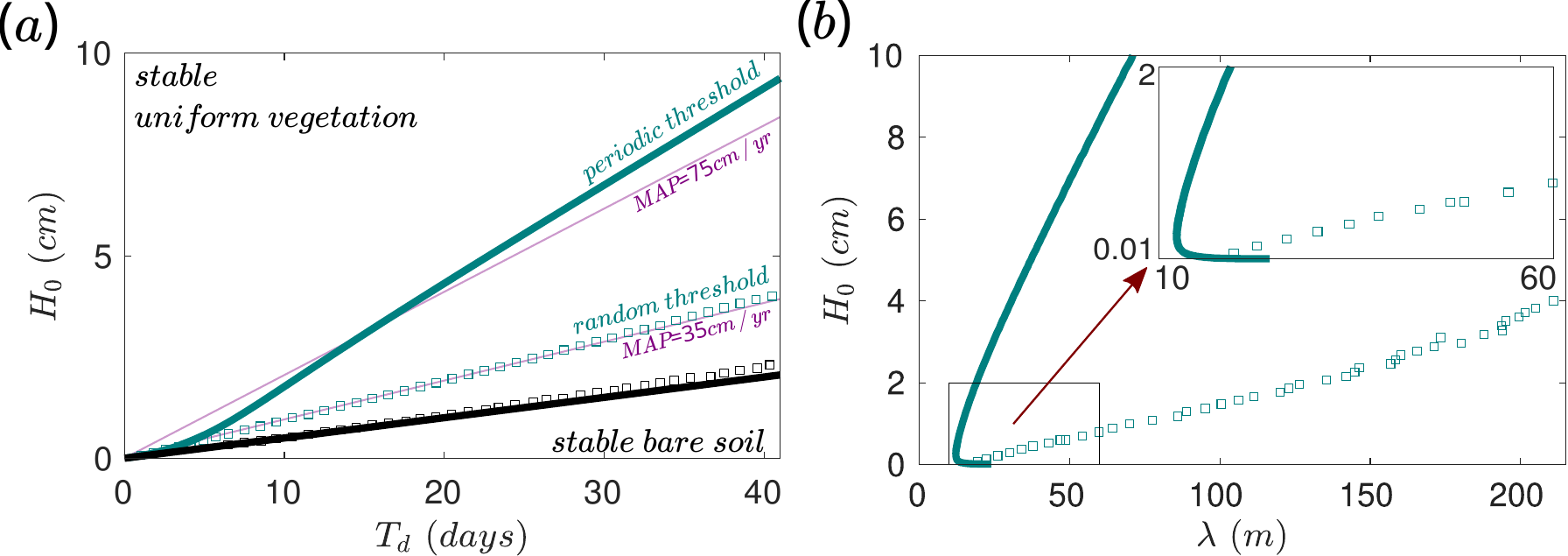}
\caption{
	(a) Thresholds for pattern forming instability in the flow-kick parameter space under both random (unfilled squares) and periodic (solid line) rainfall are shown in turquoise.  The analogous thresholds for instability of the bare soil state from Figure~\ref{fig:uniformthreshold} are included in black for comparison.  Lines of constant $\MAP=75\,cm/yr$ and $35\,cm/yr$ are also shown with thin purple. The $35\, cm/yr$ line appears at the random threshold.   (b) The wavelengths of the most unstable mode along the pattern-forming instability threshold from panel (a) are shown for both random (unfilled squares) and periodic (solid line) rainfall. 
}
\label{fig:patt:dth0curve}
\end{figure}
Before turning to numerical simulation in Section~\ref{sec:sim}, we first summarize the impact of rainfall variability on stability thresholds in the `flow-kick' space~\cite{meyer2018quantifying}. 
Figure~\ref{fig:patt:dth0curve}(a) shows the pattern-forming instability computed in Section~\ref{sec:pattern:instability} along with the instability of the bare soil state computed in Section~\ref{sec:transcrit} under both periodic and random rainfall. 
The solid lines indicate the threshold for instability of uniform vegetation state (turquoise) and the bare soil state (black) for periodic rainfall with storm depth $H_0$ and interstorm wait time $T_d$. Above the pattern instability threshold, the uniform vegetation state is stable and, below the bare soil stability threshold, i.e the transcritical point discussed in Sec.~\ref{sec:transcrit}, the bare soil state is stable.  The uniform vegetation state  becomes unstable (almost immediately) above the transcritical point from which it emerges. The region between the two solid curves, therefore, indicates where a uniform vegetation state exists and is unstable to spatially periodic perturbations under periodic rainfall.  
The squares in Figure~\ref{fig:patt:dth0curve}(a) indicate the location of the corresponding curve under random rainfall with storm depths drawn from a exponential distribution with mean $H_0$ and storm arrivals modeled by a Poisson point process with mean wait time $T_d$.  
As seen from Figure~\ref{fig:uniformthreshold}, the transcritical threshold (black) gets shifted up slightly -- more water input is needed for the bare soil state to give way to a uniform biomass one. The pattern instability (turquoise) is drastically shifted  downward -- the uniform vegetation state remains stable at much lower mean water input levels. The result is an overall decrease of the region in the rainfall parameter space over which patterns grow through small perturbation of the uniform vegetation state.  This in itself does not necessarily imply reduced ecosystem resilience when rainfall variability is taken into account, since bifurcations may be subcritical leading to parameter regions with at least two stable states. 
However, our numerical investigation of fully nonlinear patterns, described in Sec.~\ref{sec:sim},  further supports a conclusion that rainfall variability, absent from the idealized periodic rainfall models, inhibits the ability of patterns to rescue the ecosystems from collapse under aridity stress. Interestingly, the random threshold in Figure~\ref{fig:patt:dth0curve}(a) aligns well with a line of constant $MAP$ (purpler dotted line), indicating that the pattern forming instability becomes largely insensitive to how the rainfall is input for the random case.  This is not true for the periodic case, as can be seen by the fact that the periodic threshold clearly crosses the line of constant $MAP=75\,cm/yr$.   

Figure~\ref{fig:patt:dth0curve}(b) shows the dependence of the perturbation wavenumber on the mean storm depth along the pattern instability curve for both periodic (solid line) and random (squares) rainfall.   The general trend in both cases is a decrease in wavenumber, corresponding to an increase in wavelength, as the storm depth increases.  We note in the inset, however, a sharp increase in wavenumber with periodic rainfall for very small storm depths.  
While this small, frequent rainstorm regime is challenging  to explore numerically, these results suggest the potential for a transition in the dominant processes underlying pattern formation as the model approaches continuous, constant rainfall.  The constant-forcing pulsed-precipitation model given by equation~\eqref{eq:slow:nondim} with a constant rainfall parameter in~\eqref{eq:slow:nondim:w} does not exhibit a pattern forming instability. 
Exactly how the pattern-forming instability disappears in the limit of constant rainfall is perhaps interesting in light of recent work by Meyer et al.~\cite{meyer2024continuation}, 
which examines the impact on bifurcations for ODE flow-kick systems in that limit.

\section{Numerical Simulation} 
\label{sec:sim}

With an understanding of the impact of rainfall variability on the linear stability of the uniform vegetation state,  as detailed in Sec.~\ref{sec:pattern}, we now explore the dynamics of fully nonlinear patterns through numerical simulation of the flow-kick model described in Sec.~\ref{sec:kickflow}.  Our simulation results, with both periodic and random rainfall, confirm the linear predictions about the location of the pattern-forming instability in parameter space. For the parameter sets used, they also suggest that patterns form through a supercritical bifurcation as rainfall levels decrease.  We begin with an overview of the main results about the impact of variable rainfall in Section~\ref{sec:sim:overview}. 
Further details about the ramped rainfall simulations 
that allow us to draw these conclusions are provided in Section~\ref{sec:sim:ramp}.  Finally, in Section~\ref{sec:sim:collapse},  we explore collapse of vegetation patterns to the bare soil in the bistable regime associated with low water input.

\subsection{Overview of Main Results}
\label{sec:sim:overview}
The linear stability analysis of Section~\ref{sec:pattern} indicates  a significant decrease in the interval of instability of the uniform vegetation state under random rainfall, and in this section we provide an overview of results from numerical simulation that suggests a corresponding decrease in the interval of existence for stable patterns. 

\begin{figure}[h]
\centering
\includegraphics[width=\textwidth]{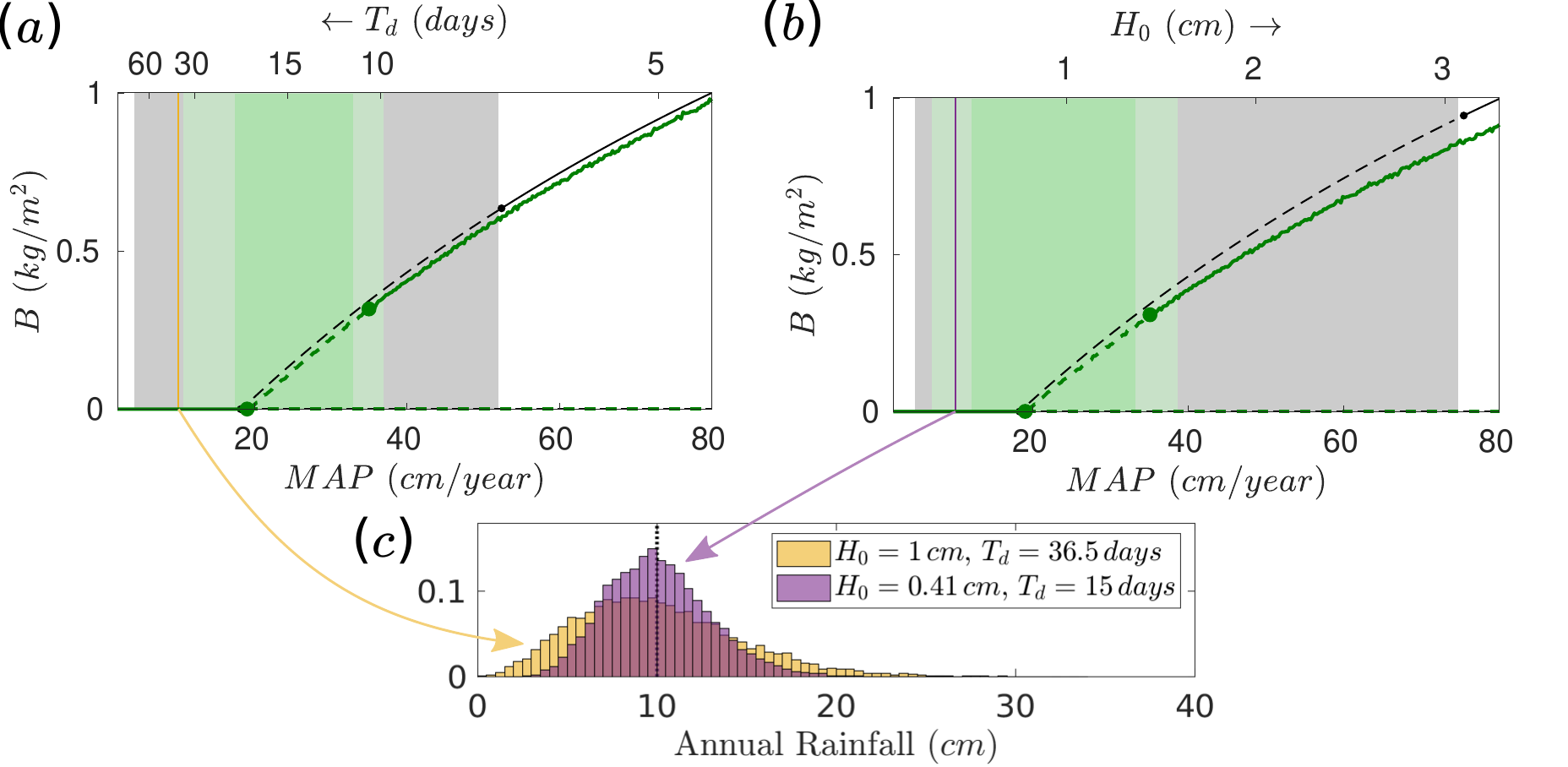}
\caption{ Mean biomass level prior to storm for uniform states with random rainfall is shown in green as a function of Mean Annual Precipitation (\MAP) where (a) mean storm depth is fixed at $H_0=1 \, cm$ while mean time $T_d$ between storm varies, and (b) mean storm depth $H_0$ varies while mean time between storms is fixed at $T_d=15\, days$. Linearly stable/unstable spatially uniform states are shown with green solid/dashed lines, and the analogous results with periodic rainfall are shown in black for comparison (see also Figure~\ref{fig:patterns:mapk}). 
	Patterns were obtained by numerical simulation with random rainfall within the green shaded \MAP\ interval, where dark green indicates that patterns were obtained in all trials while light green indicates patterns were obtained in only some fraction of trials. The \MAP\ interval in which patterns are obtained from simulation with periodic rainfall is shown with gray shading for comparison.   The mean biomass of the uniform state is lower and the parameter interval over which patterns are obtained in simulations is narrower for random rainfall. (c) Distribution of annual rainfall at $MAP=10\, cm/year$ for rainfall parameters from panel (a) with $H_0=1\,cm$ and $T_d=36.5\,days$, and panel (b) with $H_0=0.411\,cm$ and $T_d=15\, days$  are shown in yellow and purple, respectively. 
}
\label{fig:sim:intro}
\end{figure}

Figure~\ref{fig:sim:intro} shows the mean biomass level, recorded at the storm kick arrival times,  for the uniform vegetation state with random rainfall as a function of \MAP\ in green, with stability/instability indicated by solid/dashed line.  The black line represents the analogous biomass level with periodic rainfall.  We see that one effect of the random rainfall is to lower the biomass level of the uniform vegetation state at a given \MAP\ value.  The biomass level is also no longer insensitive to how \MAP\ is changed, as it is for periodic rainfall. (The black curves are identical in panels (a) and (b).)  
Indeed, decrease in biomass is more pronounced for the same \MAP\ in panel (b) with increasing storm depth than in panel (a) with decreasing interstorm dry period.  

In addition to a decrease in biomass, we see a decrease in the \MAP\ range where the uniform vegetation state is unstable. The onset of instability of the uniform vegetation state drops significantly, to $\MAP=34.8\,cm/year$ in {\it both} Figures~\ref{fig:sim:intro}(a) and (b).   In fact, with random rainfall, our numerical calculations suggest that the onset of the pattern forming instability always occurs at this \MAP, independent of the particular choice of $H_0$ and $T_d$ associated with it. The decrease in the upper instability threshold is in addition to the slight increase in the \MAP\ value of the transcritical point noted in Section~\ref{sec:transcrit}. There is also a decrease in the interval where patterns are observed in numerical simulation that goes beyond the decreased range of instability. The gray shaded regions in Figure~\ref{fig:sim:intro} indicate the parameter interval over which patterns are obtained from simulations described in Section~\ref{sec:sim:ramp} with periodic rainfall.  The green shading within the gray region shows where patterns are obtained with random rainfall.  Here, dark green indicates that patterns were obtained in all 50 of the relevant  
trials for the given rainfall parameters, while only some fraction of these simulations produced patterns in the light green shaded region on either edge.   

The simulation results indicate that there is little to no region where stable patterns co-exist with a stable uniform vegetation state near the pattern-forming instability in both random and periodic rainfall cases, suggesting a supercritical bifurcation to patterns.  At low water input, on the other hand, we see a region of co-existence between stable patterns and stable bare soil.  The introduction of randomness to the rainfall model decreases this region with the disappearance of patterns occurring at a higher water input level.   The impact is greater in Figure~\ref{fig:sim:intro}(a) where the storms are more intense and less frequent than in Figure~\ref{fig:sim:intro}(b) where low \MAP\ values are obtained by decreasing mean storm depth. This suggests that increased variability in rainfall may negatively impact the resilience of patterns at low precipitation levels.  Figure~\ref{fig:sim:intro}(c) shows  the distribution of annual rainfalls at $MAP=10\; cm/year$      
with $H_0=1\, cm$ (and $T_d=36.5\, days$), corresponding to panel (a), in yellow and with  $T_d=15\, days$ (and $H_d=0.411\,cm$), corresponding to panel (b), in magenta. Indeed, there is a larger variance in the distribution associated with panel (a), where the patterns are not observed in any of the simulations at $\MAP=10\,cm/year$.  We note that, with random rainfall, we expect a finite lifetime for the patterns before collapsing in the parameter regime below the transcritical point, where they co-exist with a stable bare soil state. 
Section~\ref{sec:sim:collapse} further explores how rainfall variability impacts the collapse of patterns to the bare soil in this low water input regime.

\subsection{Numerical Experiments with Slowly-Varying Rainfall Characteristics}
\label{sec:sim:ramp}

\tikzstyle{bnode} = [rectangle, draw, fill=green!20,   minimum height=1em, text width=4em,align=center]
\tikzstyle{rnode} = [rectangle, draw, fill=green!20,   minimum height=1em, text width=4em,align=center]
\tikzstyle{outblock} =[circle, draw, fill=black!20, minimum height=1em]
\tikzstyle{inblock} = [circle, draw, fill=black!20, minimum height=1em]
\tikzstyle{simblock} = [rectangle,
draw, fill=blue!20, text centered, rounded corners, minimum height=2em, text width=4em,align=center]
\tikzstyle{stepblock} =[rectangle,
draw,  fill=red!20, text centered, rounded corners, minimum height=2em,text width=4em,align=center]
%
\tikzstyle{simline} = [draw,arrows = {-Stealth[length=10pt, inset=2pt]}]
\tikzstyle{stepline} = [draw, dotted, -latex']
\tikzstyle{ioline} = [draw,dashed, -latex']

\begin{figure}
(a)\hspace{6cm} (b) \\
\begin{minipage}{.40\textwidth}
	\resizebox{\textwidth}{!}{
		\begin{tikzpicture}[node distance = 1cm, auto]
			\node (t0)  [] {};
			\node (sim1) [simblock, above=of t0 ] {Simulate for 50 years};
			\node (t1o)  [above=of sim1] {};
			\node (bout1) [bnode, right=of t1o] {Output annually averaged $(W,B)$};
			\node (t1i)  [right=of bout1 ] {};
			\node (step1) [stepblock, below=of t1i] {Step $MAP$ by 0.1 $cm/year$};
			\node (rand1) [rnode, right=of t0] {add spatial noise to $(W,B)$};
			\node (t2o)  [right=of  rand1 ] {};

			\path[simline] (sim1.north) -- (bout1.west);
			\path[simline] (bout1.east)--(step1.north);
			\path[simline] (step1.south) -- (rand1.east);
			\path[simline] (rand1.west)--(sim1.south);
		\end{tikzpicture}
	}
\end{minipage}
\hfill
\begin{minipage}{.58\textwidth}
	\resizebox{\textwidth}{!}{
		\begin{tikzpicture}[node distance = 2cm, auto]
			\node (uniform) [bnode,text width=5em] {Uniform vegetation};
			\node (bare) [bnode, right=of uniform]{Bare soil};
			\node (kpatt) [bnode, right=of bare,text width=6em]{Pattern with $K\; bands/km$};
			\node (patt) [rnode,above=of kpatt] {Pattern};
			\node (nopatt) [rnode,above=of $(uniform.north)!0.5!(bare.north)$] {No pattern};
			\node (bout1) [bnode,above=of $(nopatt.north)!0.5!(patt.north)$, text width=10em] { Annually averaged $B$: \\  $\Delta B=B_{\max}-B_{\min}$ };

			
			\path[simline] (bout1)--(patt)
			node () [midway,above,sloped] {$\Delta B>B_{\epsilon}$};
			\path[simline] (bout1)--(nopatt)
			node () [midway,above,sloped]  {$\Delta B\le B_{\epsilon}$};
			\path[simline] (nopatt)--(uniform)
			node() [midway,above,sloped] {$B_{\max}>B_{\epsilon}$};
			\path[simline] (nopatt)--(bare)
			node() [midway,above,sloped] {$B_{\max}\le B_{\epsilon}$};
			
			\path[simline] (patt)--(kpatt)
			node() [midway,left,text width=4em, align=right] {fit to constant biomass bands};
			
		\end{tikzpicture}
	}
\end{minipage}
\caption{(a) Schematic of simulation protocol for ramped numerical experiments. (b) Decision tree for classifying state based on annually averaged biomass $B$ in ramped experiments. 
}
\label{fig:sim:schematic}
\end{figure}
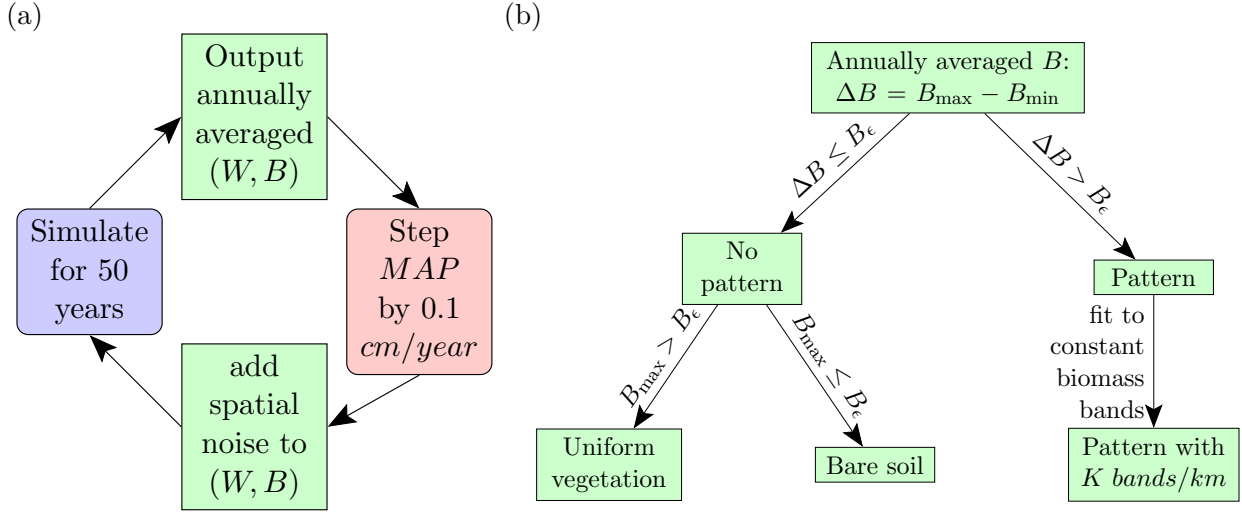
In this section, we describe details of the numerical experiments used to infer the gray and green shaded regions of Figure~\ref{fig:sim:intro} for  periodic and random rainfall, respectively.  We find agreement between the appearance of patterns in simulations and onset of the pattern forming instability computed in Section~\ref{sec:pattern:instability} with both periodic and random rainfall. Perhaps surprisingly, however, we find that the linear stability predictions about pattern wavenumber align reasonably well with our simulation results in the random rainfall case, which is definitely not the case for periodic rainfall. 

We probe the region of existence for stable patterns for both periodic and random rainfall by slowly ramping \MAP\ up and down.  We do so in two ways: (a) by varying the interstorm time $T_d$ at fixed storm depth $H_0=1\,cm$, and (b) by varying the storm depth $H_0$ at fixed interstorm time $T_d=15\, days$.  The ramping procedure is summarized by the schematic in Figure~\ref{fig:sim:schematic}(a).  We step up and down in \MAP\ at increments of 0.1 $cm/year$ and, at each step, we add small uniform random  noise to the state before simulating with fixed rainfall parameters for 50 $years$.  We use the annually-averaged biomass profile output, from the last year of this simulation period, to classify the state of the system according to the decision tree shown in Figure~\ref{fig:sim:schematic}(b).  Specifically, we first determine whether or not the state is spatially uniform by comparing the maximal difference in biomass across the domain, $\Delta B=B_{\max}-B_{\min}$, to a threshold value  $B_{\epsilon}=0.02\, kg/m^2$.  
If $\Delta B > B_\epsilon$, we consider the system to be in a spatially non-uniform state, and use a least-squares fit of the biomass profile to an idealized piecewise-constant distribution that can take on only two values, either $B=0$ or $B=B_b$. 
We use the length of each the segment in the piecewise constant distribution and the nonzero biomass level $B_b$ as fitting parameters. We find the mean wavenumber $K$ for the pattern by counting the number of $B_b$ segments, which we take to be vegetation bands, on the $1\,km$ domain. 
Otherwise, we classify the state to be either uniform vegetation or bare soil, depending on whether or not the the maximum biomass $B_{\max}$ is greater or less than the threshold $B_{\epsilon}$.   If the system reaches the bare soil state, we perturb by adding a random (positive) noise with 1\% of $B_\epsilon$ instead of 1\% of the current biomass level before stepping in $MAP$ for the next simulation period.

\paragraph{Periodic Rainfall}

\begin{figure}[ht!]
\centering
\includegraphics[width=\textwidth]{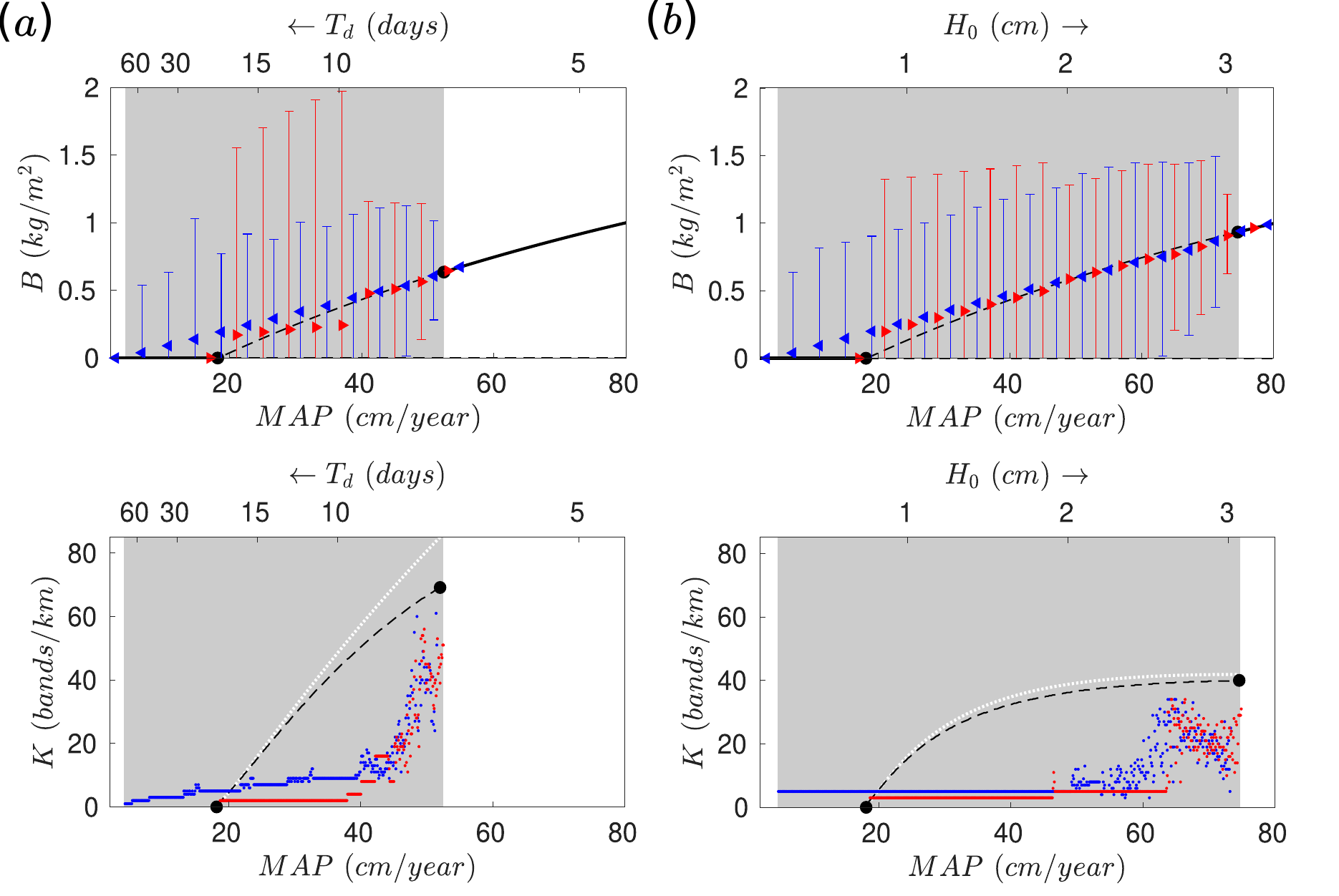}
\caption{\textbf{Ramped Simulations with Periodic Rainfall.} Biomass levels (upper panels) and mean wavenumbers (lower panels) for patterns from simulations in which \MAP\ is decreased (blue) and increased (red) by (a) changing the dry period $T_d$ between storms with fixed $H_0=1\,cm$ and by (b) changing the storm depth $H_0$ with fixed $T_d=15\,days$. The \MAP\ range over which simulations produce patterns is shaded gray.   In the upper panels, the biomass levels of the uniform state is shown in black with linear stability and instability indicated by solid and dashed lines. The range of biomass values within the patterns is indicated by vertical bars, and the domain-averaged biomass level is marked with triangles.  In the lower panels, the black dashed line indicates the wavenumber of the fastest growing unstable mode from the linear analysis,  and the white dotted line shows the wavenumber associated with $k_1^*=\pi/\ell_0$ resonance.  Note that only a subsample of the simulation points are shown in the upper panels in order to maintain readability.     
}
\label{fig:patterns:simP}
\end{figure}

We begin with simulations under periodic rainfall, which will serve as our baseline for comparisons with the random rainfall model.  The results described here are summarized by the gray shaded regions in Figure~\ref{fig:sim:intro} that indicate \MAP\ ranges over which  patterns are observed in the ramp experiments with protocols summarized in Figure~\ref{fig:sim:schematic}. 

The upper panels of Figure~\ref{fig:patterns:simP} show the biomass levels of the patterns and uniform vegetation states observed in these ramped numerical experiments with periodic rainfall when \MAP\ is varied by (a) ramping the dry period $T_d$ between storms with fixed $H_0=1\,cm$ and (b) ramping the storm depth $H_0$ with fixed $T_d=15\,days$.  In both cases, The appearance of patterns aligns well with the instability point of the uniform state  
computed in Section~\ref{sec:pattern:instability}.  There is a slight difference  in the \MAP\ level ($\sim 0.5\, cm/year$) where the patterns first appear for simulations with increasing and decreasing  \MAP\ (not visible in Figure~\ref{fig:patterns:simP}), but this is likely due to bifurcation delay. 
Indeed, ramped simulations near this instability point with rate of change of \MAP\ decreased by a factor of eight lead to a factor of two decrease in the size of the discrepancy and improved agreement between the simulations and linear predictions.   

\begin{figure}
\centering
\includegraphics[width=\linewidth]{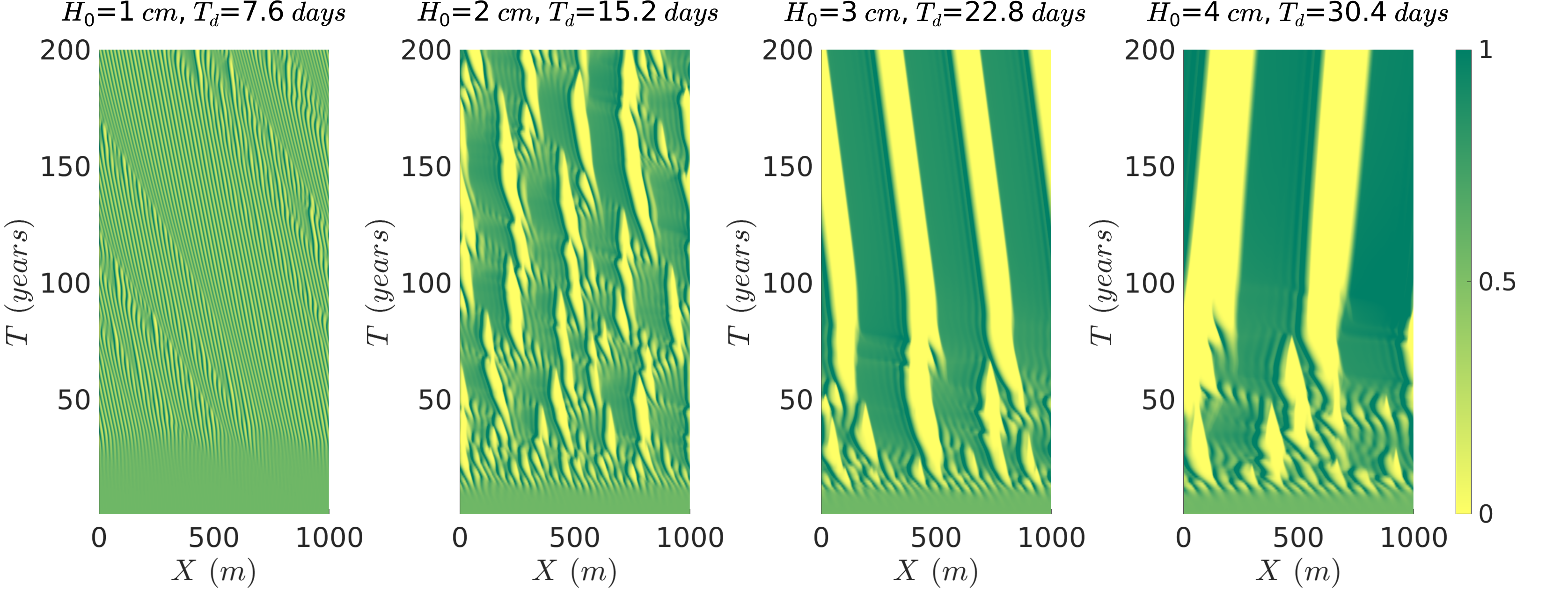}
\caption{  Spacetime plots of annually-averaged biomass distribution from 200 year simulations on a 1 $km$ domain with periodic rainfall. The mean annual precipitation is set to $MAP=48\, cm/year$ for all cases, but different values of the storm depth $H_0$ and interstorm time $T_d$ are used in each.}
\label{fig:sim:perBxt}
\end{figure}

While the pattern forming instability computed in Section~\ref{sec:pattern} agrees with the appearance of patterns in simulations, the wavenumber of the simulated patterns does not match linear stability predictions for the fastest growing mode.  The lower panels of Figure~\ref{fig:patterns:simP} compare the mean wavenumber of the patterns (red and blue dots) that appear in the year prior to each step in $MAP$ during simulations to linear stability predictions (black dashed line) for the fastest growing mode. 
While significantly higher than those observed in simulation, the wavenumber from linear stability predictions do match well with predictions based on the distance surface water travels following a storm, assuming biomass levels of the uniform vegetation state (dotted white line).   As noted in~\cite{gandhi2023pulsed}, numerical computation of the distance surface water travels over the patterned state is greater than the distance it would travel over uniform vegetation; the former provides useful information about the band spacing for nonlinear patterns.  

The patterns observed in the \MAP\ interval starting just below the onset of instability of the uniform vegetation state also have a qualitatively different character than simple traveling wave solutions of the type shown in Figure~\ref{fig:nm:example}.    
These states exhibit complicated spatiotemporal dynamics with fluctuating number of bands and biomass traveling both uphill and downhill. This behavior is reflected in the fluctuations in wavenumber observed above about $\MAP=40\,cm/year$ in the lower left panel of Figure~\ref{fig:patterns:simP}, and above about $\MAP=47\, cm/year$ in the lower right panel. Figure~\ref{fig:sim:perBxt} illustrates a variety of different patterns that can be obtained at a fixed mean annual precipitation of $MAP=48\, cm/year$, and different combinations of storm depth $H_0$ and inter-storm time interval $T_d$. At $H_0=1\,cm$ (and $T_d=7.6\, days$), the pattern has a mean wavenumber of around $K=50\, bands/km$ and undergoes long-wave instabilities.   For simulations where the storms become more intense and less frequent, the pattern has fewer bands.  For storm depths above about $H_0=3\, cm$, the simulations produce traveling wave patterns with the bands migrating uphill or downhill.  This sensitivity of the travel speed to small parameter changes disappears when random rainfall is introduced~\cite{gandhi2023pulsed}.        

In the ramped numerical experiments of Figure~\ref{fig:patterns:simP}, 
simple traveling wave patterns, like those shown in the right panels of Figure~\ref{fig:sim:perBxt}, are observed for low enough \MAP\ levels.  The transition from the complex spatiotemporal patterns that appear at onset occurs when the periodic storms are (a) infrequent enough at fixed $H_0=1\, cm$, or (b) mild enough at fixed $T_d=15\, days$.    While there are occasional transitions in the number of bands, often associated with jumps in the peak and/or mean level of biomass, the pattern wavenumber remains consistent over large intervals in \MAP\ until they disappear at low $MAP$ levels.  We note that patterns in simulations with decreasing \MAP\ persist far below the transcritical bifurcation point  
of the uniform states, to about $4\, cm/year$ in Figure~\ref{fig:patterns:simP}(a) and to about $5\, cm/year$ in Figure~\ref{fig:patterns:simP}(b). This results in a region where patterns and a bare soil state stably co-exist.  Overall, we see a gradual 
decrease in mean biomass levels as \MAP\ decreases.  We further note that the rate of decrease of biomass levels, with decreasing \MAP, for the patterned state is slower than the rate of decrease for the uniform vegetation state. This captures the way that patterns are more productive than uniform biomass below the pattern-forming instability.

\paragraph{Random Rainfall}

\begin{figure}[ht!]
\centering
\includegraphics[width=0.95\textwidth]{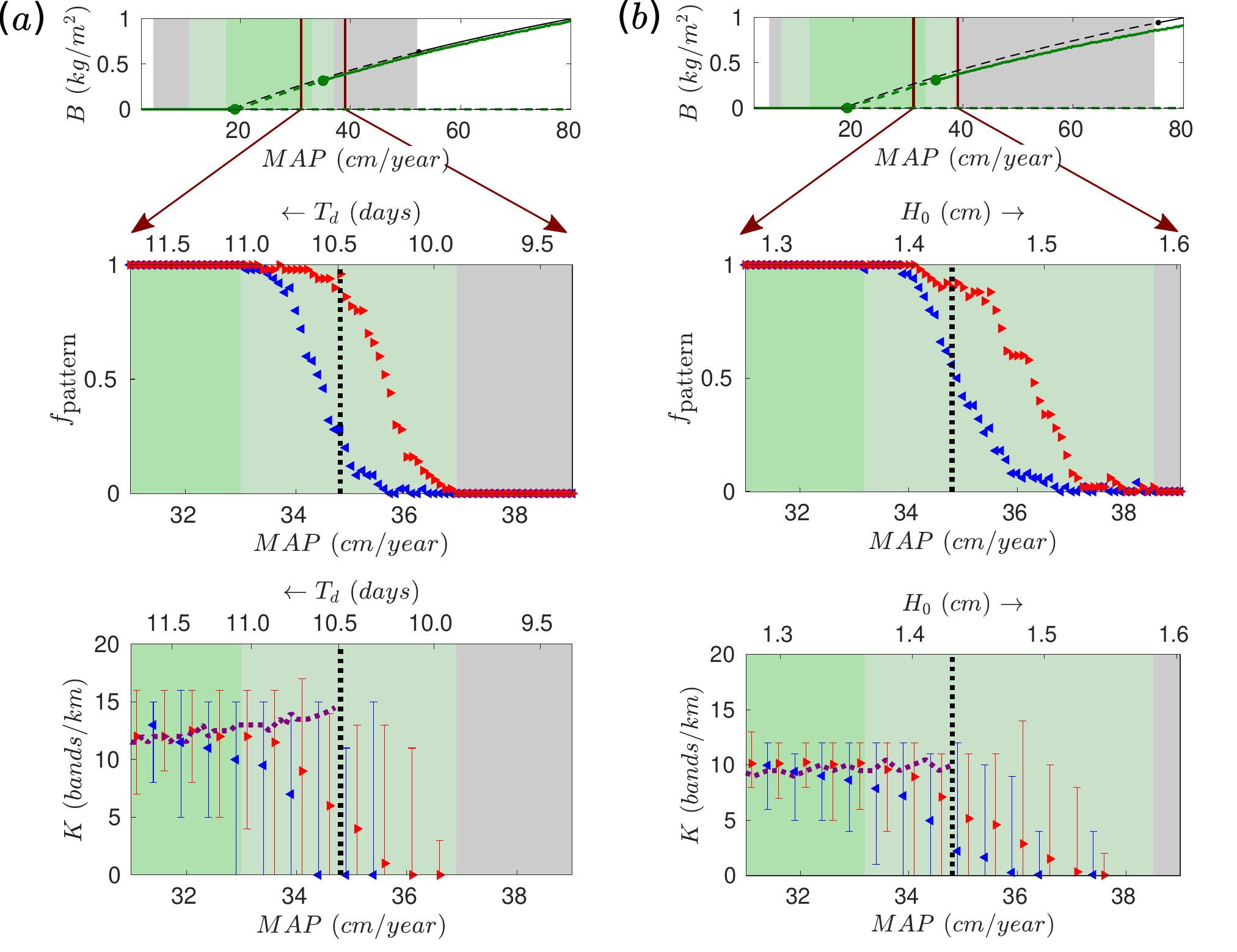}
\caption{\textbf{Ramped simulations  near upper instability with random rainfall.}  
	Simulations results from 50 trials in which \MAP\ is decreased (blue) and increased (red)  near the  pattern forming instability (vertical black dotted line).  \MAP\ is varied by (a) changing the dry period $T_d$ between storms with fixed $H_0=1\,cm$ and by (b) changing the storm depth $H_0$ with fixed $T_d=15\,days$.   All trials with \textit{increasing} \MAP\ result in spatially non-uniform states within the dark green shaded region, while some fraction of these trials produce non-uniform states within the light green region.  The upper panels show the fraction of trials in which the state is classified as non-uniform at the end of the 50-year period.  In the lower panels, the range of mean pattern wavenumbers across trials is indicated by vertical bars, and median pattern wavenumber is marked by triangles.  The wavenumber of the fastest growing mode based on linear stability analysis is indicated by the purple dotted line.   Note that only a subsample of the simulation points are shown in the lower panels in order to maintain readability.  
}
\label{fig:patterns:simR:upper}
\end{figure}

\begin{figure}[ht!]
\centering
\includegraphics[width=0.95\textwidth]{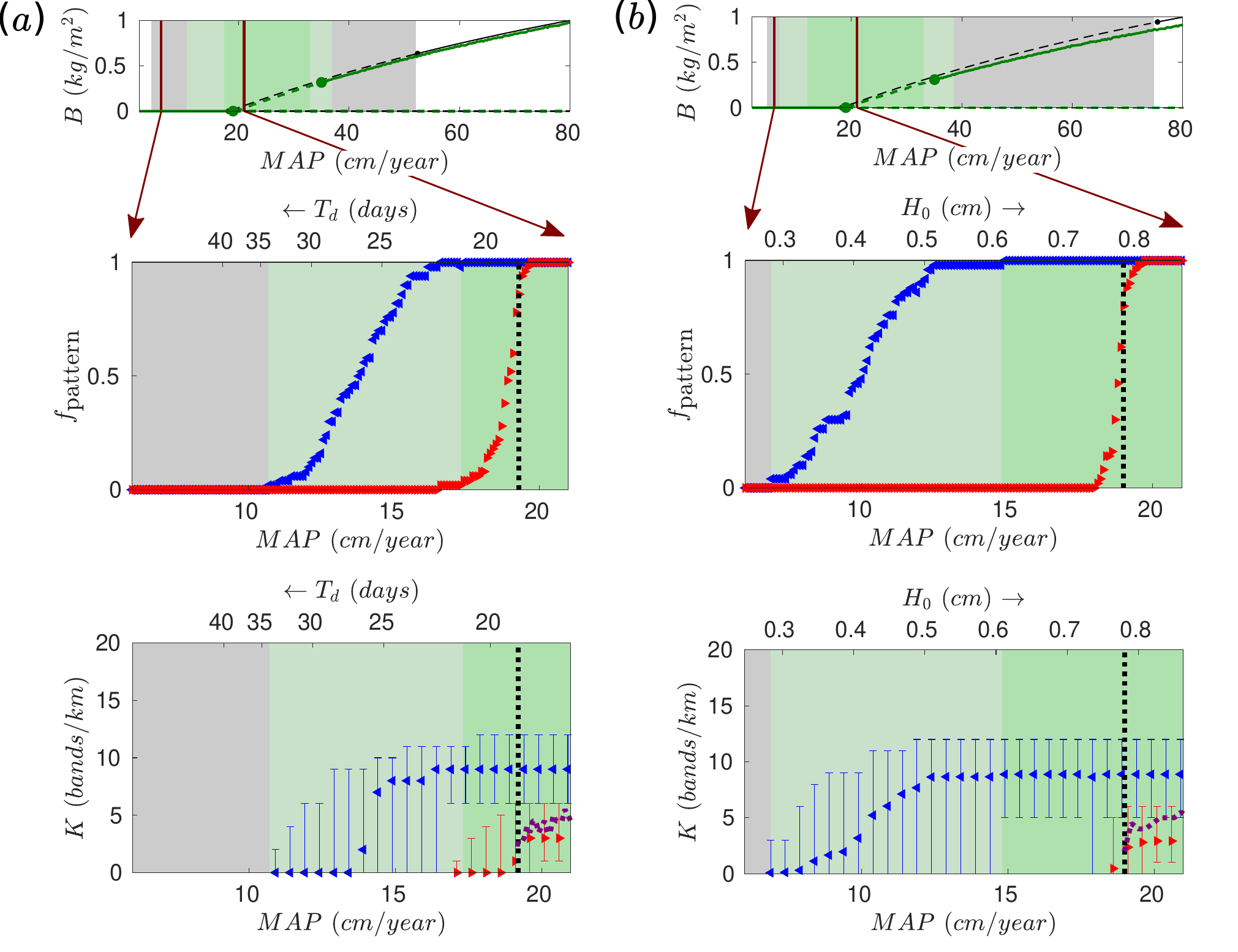}
\caption{\textbf{Ramped simulations  near lower instability with random rainfall.}       Simulation results from 50 trials in which \MAP\ is decreased (blue) and increased (red)  near the  transcritical point (vertical black dotted line).  \MAP\ is varied by (a) changing the dry period $T_d$ between storms with fixed $H_0=1\,cm$ and by (b) changing the storm depth $H_0$ with fixed $T_d=15\,days$.   All simulation trials with \textit{decreasing} random rainfall result in spatial patterns within the dark green shaded region, while some fraction of the trials produce patterned states within the light green region.  The upper panels show the fraction of trials in which the state is classified as a pattern at the end of the 50-year period.  In the lower panels, the range of pattern wavenumbers across trials is indicated by vertical bars, and the median wavenumber by triangles.  The wavenumber of the fastest growing mode is indicated by the purple dotted line, which necessarily terminates at the transcritical bifurcation.   Note that only a subsample of the simulation points are shown in the lower panels in order to maintain readability.
}
\label{fig:patterns:simR:lower}
\end{figure}

We follow the same simulation protocol summarized by Figure~\ref{fig:sim:schematic} with random rainfall. 
We carry out a small number of 5 trails 
in which we ramp over the the entire \MAP\ range and then run 50 trials of ramping up and down near the observed transition points. Figure~\ref{fig:patterns:simR:upper} shows the results from these simulations near the pattern-forming instability where the uniform state first loses stability, which occurs at (a) mean interstorm time of $T_d=10.5\; days$ with fixed mean storm depth $H_0=1\; cm$ and (b) mean storm depth of $H_0=1.42\;cm$ with fixed mean dry periods between storms of $T_d=15\; days$.   
These instability points are marked by a vertical black dotted line. The top panels show the fraction of the 50 trials leading to a patterned state, denoted $f_{\rm pattern}$, at each  \MAP\ value as it is increased (red) and decreased (blue).  The dark green shading here and in Figure~\ref{fig:sim:intro} indicates where all 50 simulations, for increasing (above lower threshold) and decreasing (below upper threshold) \MAP, exhibit patterns.   In the light green shaded region, at least one simulation exhibits patterns and at least one does not.  The transition where patterns begin to appear and disappear aligns with the instability points predicted by the linear stability analysis in Section~\ref{sec:pattern}.  There is a small discrepancy in when the patterns begin to appear with decreasing \MAP\ versus when they begin to disappear with increasing \MAP, and this is likely a delay effect analogous to the bifurcation-delay observed with periodic rainfall.  (We describe the impact of changing the rate of the ramp that decreases \MAP\ in Section~\ref{sec:sim:collapse}.)

In contrast to the case with periodic rainfall, we see agreement between the pattern wavenumbers observed in simulation and predictions from linear stability for the case of random rainfall.  The lower panels in Figure~\ref{fig:patterns:simR:upper} show the median wavenumber of the patterns from the simulations with  triangles and the vertical bars indicate the maximum and minimum wavenumber obtained.  
The predictions of the fastest growing mode based on linear stability analysis of Section~\ref{sec:pattern:instability}, shown with purple dots, falls within the range of wavenumbers obeserved in simulations.  In particular, we typically see a higher number of bands per kilometer in simulations near onset with fixed $H_0=1\, cm$ than near onset with fixed $T_d=15\, days$, just as predicted by our linear stability analysis.  This is also consistent with the prediction, based on surface distance traveled by water during storms, that the wavelength of the preferred pattern is proportionately  
larger for more intense storms.          

We also probe the lower precipitation range with ramped numerical experiments that start sufficiently far above the transcritical point so that all simulations initialized with small random noise quickly approach a patterned state for decreasing \MAP. 
For increasing \MAP,  we start at low enough precipitation levels to ensure that all simulations remain in the bare soil state for the initial steps of the ramp up. 
Figure~\ref{fig:patterns:simR:lower} shows the resulting fraction of simulations in the patterned state and median wavenumbers as a function of \MAP\ during these ramped simulations.  Patterns begin to appear in the increasing \MAP\ simulations near the transcritical point predicted by linear stability analysis.  The  lower panels again show good agreement of wavenumbers with the linear stability predictions of the most unstable mode. 

\subsection{Collapse at low rainfall}
\label{sec:sim:collapse}
The ramped numerical experiments presented in Section~\ref{sec:sim:ramp} show  persistence of patterned states, over the decades-to-centuries timescale of these simulations, in the bistable regime below the transcritical point. However, due to the stochasticity of the model, we expect rare event transitions to the stable bare soil state can occur for any parameter set below the transcritical point.  In this section, we explore the influence of this finite lifetime for patterns in the presence of an alternate stable bare soil state.

\begin{figure}[ht!]
\centering
\includegraphics[width=\textwidth]{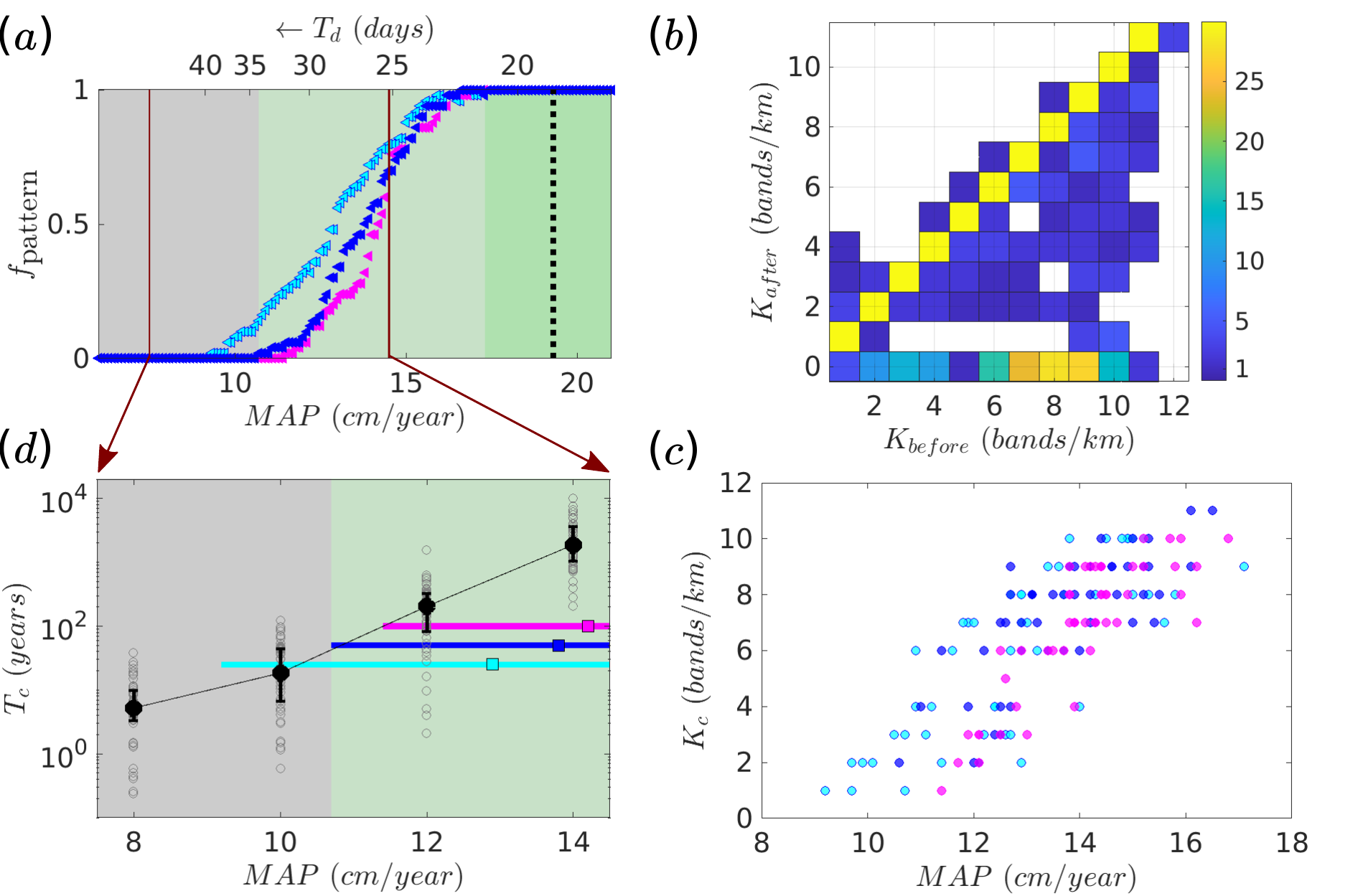}
\caption{(a) Fraction of simulations with patterns in simulations where \MAP\ is slowly decreased by changing $T_d$ at fixed $H_0=1\,cm$.  For simulations shown with blue arrows, \MAP\ is stepped down by 0.1 $cm/year$ every 50 years, as in Figure~\ref{fig:patterns:simR:lower}(a). The ramp speed is doubled (i.e. stepped down by 0.1 $cm/year$ every 25 years) for the simulations shown in cyan and halved (i.e. stepped down by 0.1 $cm/year$ every 100 years) for those shown in magenta. 
	(b) Bivariate histogram of wavenumber before and after transitions between patterned states in the ramp down simulations from panel (a).  The diagonal elements, corresponding to no transition, are colored yellow. Collapse events in which a patterned state transitions to bare soil are marked here along the $K_{after}=0$ line.  (c) For all collapse events, the wavenumber just prior to the transition to bare soil is plotted as a function of \MAP\ at which the collapse occurred.  The points are colored according to ramp rate as in panel (a).  (d) Median (solid black circles) and interquartile range (black vertical bars) of times $T_c$ to collapse in simulations initialize with a single band on a 200 $m$ domain with random rainfall. The mean storm depth is $H_0=1\; cm$ and the mean interstorm time $T_d$ is set so that $\MAP=8,\, 10,\, 12$ and 14 $cm/year$, with 50 trials carried out for each case.   The parameter range over which collapse events from panel (a) are observed for each set of trials is indicated by a horizontal line of the same color,  with a square marking where 50\% of the trails have collapsed.  
	\label{fig:sim:collapse}
}
\end{figure}

We begin by repeating the ramped numerical experiment with decreasing \MAP\ at fixed $H_0=1\, cm$  shown in Figure~\ref{fig:patterns:simR:lower}(a) with different ramp rates.  Figure~\ref{fig:sim:collapse}(a) shows the fraction of simulations that exhibit patterns from the original simulations with 0.1 $cm/year$ steps in \MAP\ every 50 years in blue triangles. The analogous results with the ramp rate doubled so that the 0.1 $cm/year$ steps occur every 25 years are shown with cyan triangles, and the results with half the  ramp rate, with steps every 100 years, are shown in magenta.   There is a slight steepening of the curves as the ramp rate is decreased but the overall parameter range over which the collapses occur remains largely unchanged.    
The system supports a large multiplicity of  patterns, and may thus take a variety of transition paths to collapse.   
The most likely mean wavenumber appearing in the simulations is $K=9\,bands/year$, independent of the ramp rate. We observe transitions between patterns of different wavenumber prior to collapse in roughly 1/3 of the simulations and, with decreasing \MAP, a large majority of these non-collapse transitions involve loss of bands. The most likely transitions observed for all three ramp rates is a loss of 1 or 2 bands, although almost all possible transitions involving a loss of bands are observed. We note that this is in contrast to standard reaction-advection-diffusion models with a slowly-varying rainfall parameter that often predict collapse through a cascade of period doublings in which half the vegetation bands are lost at each step~\cite{siteur2014beyond,rietkerk2021evasion}.  
Figure~\ref{fig:sim:collapse}(b) shows a bivariate histogram of the wavenumber before and after transitions between patterns.  Transitions from a pattern to the bare soil appear here with $K_{after}=0$.   

Collapse events, in which the pattern irrecoverably transitions to the bare soil state, occur within the light green shaded region of Figure~\ref{fig:sim:collapse}(a) for the standard ramp rate indicated with blue.  The wavenumber prior to collapse for each of these collapse events is  shown with blue dots in Figure~\ref{fig:sim:collapse}(c) as a function of the \MAP\ value where the collapse occurs during each simulation. We see that, in general, patterns with fewer bands can survive to lower \MAP\ before undergoing collapse.  This trend holds for all three ramp rates considered, as indicated by the addition of the cyan and magenta dots to the blue ones; these are obtained with half and double rates, respectively. 
We also note the relatively high wavenumbers of the patterns that collapse straight to bare soil, instead of transitioning to a pattern with fewer bands,   as \MAP\ decreases.  These transition typically occur for higher values of \MAP.     

We expect collapse events will happen eventually, even in simulations with fixed rainfall parameters, due to the stochastic nature of the rainfall model and the existence of an alternative stable bare soil state. We explore the finite lifetimes of the patterned states through numerical simulations initialized with a single band on a 200 $m$ domain, corresponding with a periodic pattern of wavenumber $K=5\, bands/km$.   Figure~\ref{fig:sim:collapse}(d) shows the time $T_c$ for the band  to collapse to the bare soil state in 50 trials for each set of rainfall parameters consisting of $H_0=1\, cm$ and different values of $T_d$ so that $MAP=8,$ 10, 12 and 14 $cm/year$ (empty gray circles).   For each set of trials, the median lifetime and interquartile range are indicated by the solid black circle and vertical bars.    
For comparison, we also include horizontal lines to indicate the \MAP\ range over which collapse events are observed in the ramped simulations. Each line represents a different ramp rate, using the same color code as in panel (a), and the square indicates where 50\% of the simulations have collapsed.  The horizontal lines are arranged along the vertical axis according to the time interval between the 0.1 $cm/year$ steps in \MAP\ for each set of ramped simulations. 
The lower end of the \MAP\ range for each of the three ramp speeds aligns closely with the trend line (black) for the mean collapse time at the given \MAP\ value. 
The collapse events observed in the ramped simulations therefore occur when the time spent at a given parameter set is on the order of the mean lifetime of a state with those same parameters.

\section{Discussion/Conclusions}
\label{sec:discussion}

Dryland ecosystems, which support about a third of the global population~\cite{middleton1992world}, are adapted to survive with limited and highly-variable rainfall.  Climate change is expected to impact the distribution of the rare rainfall events, particularly in  the frequency of extremes, leading to both increases in flash floods and `flash droughts' \cite{IPCC_2021}. These threats motivate some recent efforts to understand how the characteristics of precipitation, beyond the mean, may impact the susceptibility of a dryland ecosystem to collapse~\cite{dordorico2012hydrologic,gherardi2019effect}. 
Our particular focus is on dryland regions where pattern formation, in the form of banded vegetation, emerges in response to aridity stress.   This suggests that increased rainfall variability might be expected to increase the prevalence of patterns since it acts as an additional stressor on the ecosystem. We find the opposite effect, however, that patterns appear over narrower range of parameters. If rainfall arrives in larger, but more rare impulses,  then  vegetation bands that rely on  run-off from up-slope may be particularly at risk.

We have used a conceptual modeling framework~\cite{gandhi2023pulsed} for dryland vegetation pattern formation that explicitly incorporates details about the timing and amount of water input from individual rainstorms. These rain events take the form of  instantaneous, spatially heterogeneous kicks to the soil water as it interacts, through positive feedbacks, with a biomass field representing the vegetation.  We model the rainfall variability with a Poisson point process for the arrival of rainstorms, whose storm depth is drawn from an exponential distribution.  The model predicts that the randomness leads to (1) an overall decrease in the total (time-averaged) biomass, and (2) a reduction in the range of mean annual precipitation (\MAP) over which patterns are observed. 
The reduction occurs through both a \textit{decrease} of the upper \MAP\ value at which patterns first appear through an instability of the uniform vegetation state and an \textit{increase} in the lower \MAP\ value at which patterns disappear and only the bare soil desert state remains.  
Moreover, we find that the magnitude of these effects depends on the specific way in which the mean timing $T_d$ between storms and the mean storm depth $H_0$ lead to a particular \MAP\ level. 

The linear predictions of pattern-forming instabilities, presented in Section~\ref{sec:pattern}, rely on numerical approximation of maximal Lyaponuv exponents of the linearized system subjected to spatially heterogeneous perturbations. Specifically, this system, in the large time limit, is determined via repeated applications of random kick-flow maps~\eqref{eq:FlowKickPertMap}, where all arguments of the map, $(w_0,b_0,h_0,\tau_d)$,
are drawn from distributions depending on details of the rainfall model. In particular, while the distributions of the storm characteristics $(h_0,\tau_d)$ are prescribed, the resulting distribution of the base state for the perturbation, $(w_0,b_0)$, is only known via numerical simulations over a large number ($\sim 10^5$) of kicks. 
While negative Lyapunov exponents do not, in general, imply stability (see, e.g.~\cite{leonov2007time}), we numerically confirm in Section~\ref{sec:sim} that the onset of the pattern forming instability is well approximated by our estimated zero crossing of the maximal exponent.  Simulations demonstrate that linear calculations provide useful insights about the location of the pattern forming instability in parameter space including the wavenumber of the resulting pattern. The vegetation patterns appear as traveling waves, with fluctating width, and a clear time-averaged up-slope travel speed that can be seen in space-time plots.  (The approach we used here does not admit linear predictions about the mean travel speed since we are only able to numerically estimate the magnitude,  and not the complex phase, associated with instability.)  

The fluctuating width of the vegetation bands, observed in simulations of our random flow-kick model (e.g. Figure~\ref{fig:nm:example}(b)), suggests some new possibilities for comparing model predictions with observational studies. For example, Gowda et al.~\cite{gowda2018signatures}, in a study of vegetation bands in the Horn of Africa, were able to compare the width of more than 1000 individual vegetation bands at different time-points, decades apart, to obtain a distribution of their fractional change. It would be interesting to repeat measurements of that type using modern satellite image data obtained over shorter time periods, in conjunction with rainfall data. In that way, we might test any model predictions of fluctuations of bandwidths correlated with rainfall variability. We note that changes in bandwidth, in response to changes in rainfall, is more likely to occur than changes in the overall wavelength of the banded patterns, and may present a new opportunity for remotely monitoring the stress on the ecosystem with changing weather patterns.

Our findings that randomness decreases the range over which patterns exist in the flow-kick model studied here are not generally true for pattern forming systems.  For example, studies of so-called `stochastic Turing patterns' have shown that the inclusion of temporal stochasticity in the birth and death processes of spatial population models can actually enhance the range over which pattern formation occurs~\cite{butler2009robust,karig2018stochastic}. 
Similarly, other studies, looking into the impact of spatial noise on vegetation patterns, find that the introduction of spatial stochasticity in various model parameters may improve ecosystem resilience~\cite{yizhaq2014effects,yizhaq2017geodiversity}.
We note that, for the flow-kick model studied here, the randomness is introduced into the forcing and the local interactions remain deterministic.  This is not the case for the other studies mentioned above, and it may be interesting in future work to explore how the introduction of noise into the water-biomass interactions impacts the results presented here.

In addition to  reducing the parameter range for patterns to appear, the introduction of randomness into the flow-kick framework also qualitatively changes the structure of the instability predicted by linear theory.  
As seen in Figure~\ref{fig:patterns:mapk}, the resonances that appear with periodic rainfall are destroyed by the randomness. The complicated spatio-temporal dynamics observed from simulations in Figure~\ref{fig:sim:perBxt} disappear in favor of simple traveling wave behavior with uphill migration as well. 
Hamster et al.~\cite{hamster2024blurring} also observe that linear predictions correspond better to nonlinear states obtained from simulation when randomness is introduced.  They consider the Klausmeier model~\cite{Klausmeier1999} with spatio-temporal stochasticity in the mortality rate parameter, and find that the stochastic system tends to approach a pattern with (local) wavenumber near the most unstable mode of the deterministic system.  
The fact that predictions from linear stability seem  to more accurately reflect the fully nonlinear states observed in simulation when randomness is introduced raises interesting mathematical questions about the impact of impulsive forcing and variability on pattern formation more generally that will be explored in future work. Interestingly, the flow-kick system under random rainfall also behaves, at least qualitatively, more like a standard reaction-advection-diffusion system with constant forcing and aligns more closely to observations than the flow-kick system with periodic rainfall does.

The flow-kick model captures certain key aspects of the fast hydrology during rainstorms, such as positive biomass feedbacks and how they are impacted by storm depth.  Our impulsive treatment of storms, however, does not allow us to investigate the impacts of varying the duration of storms, which was shown by Crompton and Thompson~\cite{crompton2021sensitivity} to also be  important; storm duration, along with the storm depth, control storm intensity. The kick in our model can be modified to incorporate the storm duration by, for example, solving~\eqref{eq:fast:nondim} with a step-function rainfall rate of finite duration, instead of initializing $h$ with the entirety of the storm input at $t=0$.  This can allow for explicit inclusion of the storm duration timescale into the instantaneous kicks of the flow-kick framework, albeit at the expense of model (and computational) complexity. We note that another potentially important simplification of our fast hydrology model relative to models based on the shallow water equations (see, e.g., ~\cite{crompton2021sensitivity,crompton2019emulation,crompton2020resistance}) is that we neglect the dependence of the surface flow speed on the water height.  It will be important to compare our model predictions to more detailed models in future work.    
On the opposite end, we also do not consider slower timescales on which changes in community structure and phenotypic plasticity can occur. Adaptation  associated with these slow timescales has also been shown to impact resilience in models of dryland ecosystems~\cite{bera2021linking,bennett2023phenotypic}.

Throughout this study, we use periodic boundary conditions and therefore neglect the further detrimental effects of loss from a region through surface water runoff during rainstorms. While this choice justifies the use of a small, computationally tractable domain corresponding to a section in the middle of a larger hillslope, it does not allow us to capture water loss through runoff~\cite{siteur2014will,crompton2021sensitivity}. The work of Siteur et al.~\cite{siteur2014will} includes this effect in a model that has a similar pulsed-rainstorm structure to ours. While many details of the storm-level hydrology are simplified relative to our model, that study finds that both high and low extremes in storm intensity can be detrimental to pattern formation.  Water is unable to concentrate as much in the vegetation bands during low-intensity storms, an effect also captured by our model. During high-intensity storms, Siteur et al.~\cite{siteur2014will} find that there is significant water loss due to runoff, as the vegetation bands are unable to capture all the excess water from the bare soil region just uphill.  Periodic boundaries do not capture these additional losses, and alternate choices for boundary conditions on larger domains that more accurately capture the effects of run-off in our model will be considered in future work.

Modeling frameworks with a similar impulsive structure to that of the model studied here  have been applied in other contexts, such as savanna fires~\cite{hoyer2021impulsive}, population management and control~\cite{zeeman2018resilience,patel2024spectral}, biological pattern formation~\cite{colombo2023pulsed,colombo2024pulsed} and infectious disease~\cite{hoyer2023immuno}.  The approaches developed here may provide a path towards addressing questions about the  impact of variability in the context of impulsive systems more generally. A key feature of these systems is that the randomness \emph{cannot} be treated as small perturbations of the original deterministic behavior, thus rendering inapplicable  some of the standard tools of analysis of stochastic dynamical systems.   Our work in the context of dryland vegetation patterns highlights the need for the development of an effective time-averaged model that can capture key aspects of the random forcing, which we have demonstrated includes moments beyond the mean.

\vspace{5mm}
\paragraph{Acknowledgements.} 
 PG gratefully acknowledges support of the Simons Foundation Collaboration Grant for Mathematicians 848615.  MS gratefully acknowledges support of the  NSF-Simons National Institute for Theory and Mathematics in Biology via grants NSF DMS-2235451 and Simons Foundation MP-TMPS-00005320.  Computing resources were provided by the High Performance Research Computing (HPRC) Core Facility at Virginia Commonwealth University.



\appendix

\section{Solving for the soil moisture kick}
\label{supp:slovefast}
To determine the kick $\Delta w(x)$  we integrate~\eqref{eq:fast:nondim:w} at each location $x$ on the domain to obtain
\begin{equation}
	\label{eq:what}
	\Delta w(x)=\alpha \iota(x)
	\int_0^\infty \Theta(h(x,t))dt.
\end{equation}
This is simply the infiltration rate at $x$, which depends only on the biomass level at that location via \eqref{eq:iotanu}, times the total length of time that water is infiltrating the soil, i.e. when $h(x,t)>0$.

Gandhi et al.\cite{gandhi2023pulsed} used the method of characteristics to determine the surface water height $h(x,t)$ from \eqref{eq:fast:nondim:h}. 
In particular, the characteristic, labeled by $y$, is 
\begin{equation}
	\label{eq:txy}
	t(x;y)=\int_x^y \frac{1}{\nu(s)}ds, \quad
\end{equation}
which represents the time it takes for water to flow to $x$ from a point $y$ further uphill. Along the characteristic, $\hat{h}(x;y)\equiv h(x,t(x;y))$ satisfies an ordinary differential equation
\begin{equation}
	\label{eq:charex}
	\frac{d}{dx}\Bigl(\nu(x)\hat{h}(x;y)\Bigr)=\iota(x)
	\Theta\Bigl(\hat{h}(x;y)\Bigr), \quad  \hat{h}(y;y)=h_0.
\end{equation}
Integrating this, we obtain
\begin{equation}
	\label{eq:hhat}
	\hat{h}(x;y)\equiv h(x,t(x;y)) =\begin{cases}
		0, &{\rm if}\ x\le \xi_0(y)\\
		\frac{1}{\nu(x)}\left(\nu(y)h_0-\int_{x}^{y}\iota(s)ds \right)& {\rm if}\ \xi_0(y)<x< y,
	\end{cases}
\end{equation}
where $\xi_0(y)$ is the point on the characteristic where $\hat{h}(x,y)$ reaches zero, i.e. $\xi_0(y)$ satisfies
\begin{equation}
	\label{eq:xzeq}
	\int_{\xi_0(y)}^{y}\iota(s)ds =\nu(y)h_0.
\end{equation}

A change of variables in~\eqref{eq:what}, from $t$ to $y$, via the characteristic equation $t(x;y)$, given by \eqref{eq:charex}, yields 
\begin{equation}
	\label{eq:Omegax}
	\Delta w(x)=\alpha\iota(x)  \int_{x}^{\infty}\frac{\Theta\Bigl(\hat{h}(x;y)\Bigr)}{\nu(y)} dy,
\end{equation}
Now using~\eqref{eq:hhat} and the fact that $\nu$ is a positive function, we have
\begin{equation*}
	\Theta\Bigl(\hat{h}(x;y)\Bigr)= \begin{cases}
		0, &{\rm if}\ x>y\\
		\Theta
		\Bigl(h_0-
		\frac{1}{\nu(y)}
		\int_{x}^{y}\iota(s)ds \Bigr),& {\rm if}\ y>x.
	\end{cases}
\end{equation*}
The final expression for the kick $\Delta w(x)$, with further details in~\cite{gandhi2023pulsed}, is
\begin{equation}
	\label{eq:Omegax0}
	\Delta w(x)=\alpha\iota(x)
	\int_{x}^{y_\ell(x)}\frac{\Theta\Bigl(h_0-\hat{h}(y;y_\ell(x))\Bigr)}{\nu(y)} dy,
\end{equation}
where 
\begin{equation}
	\label{eq:hhat0}
	\hat{h}(y;y_\ell(x))=\frac{1}{\nu(y)}\int_x^y\iota(s) ds.    
\end{equation}
This integral  represents the height of the surface water at a point $y$ along a characteristic that starts at $x$ with $h=0$ and follows it backward in time to the largest $y$, denoted  $y_\ell(x)$, where $h=h_0$. 

Here we provide a simple example of computing the kick $\Delta w(x)$ for  $h_0=1$, and when the biomass is given by 
\begin{equation*}
	\hat{b}(x)=2(1+\cos(4x)), 
\end{equation*}
which is $\pi/2$-periodic. (In dimensioned units, this corresponds to a periodicity of $\sim 110 m$, with a peak biomass level of $4Q=0.4\ kg/m^2$, for the parameters of Table~\ref{tab:dim}.)

\begin{figure}
	\includegraphics[width=15cm]{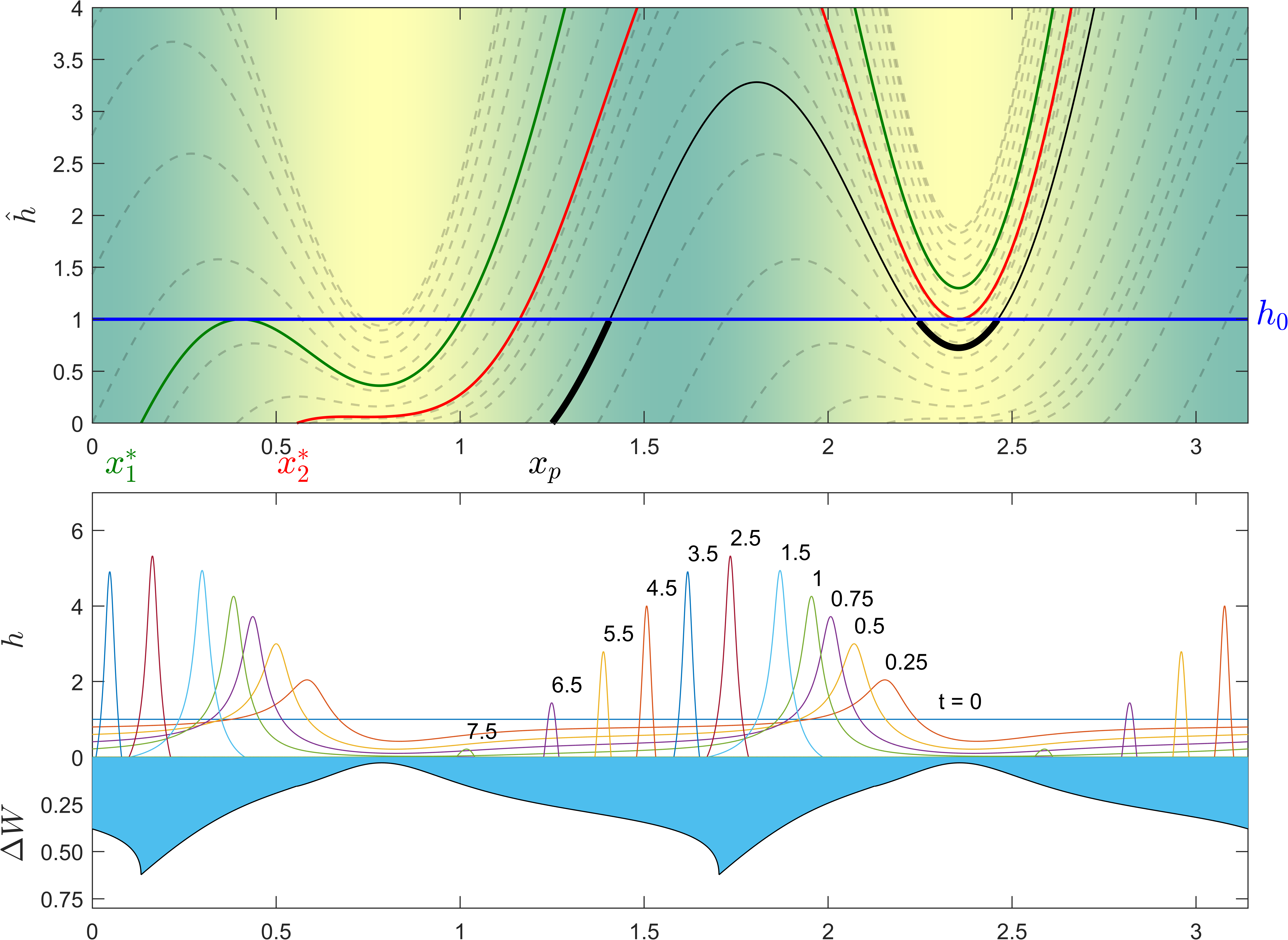}
	\centering
	\caption{Computing the kick $\Delta w(x)$ for storm depth $h_0=1$ and $\hat{b}(x)=2(1+\cos(4x))$, indicated respectively by the blue line and green shading in the top panel. (top) Characteristics $\hat {h}(x;y)$ from \eqref{eq:hhat0}. For $x\in[x_1^*,x_2^*]$ we see there is only one piece where $\hat{h}\le h_0$, so the integral in \eqref{eq:Omegax0} is over a single interval, where for $x\in[0,\pi/2]\setminus[x_1^*,x_2^*]$, there are two such intervals. As an example, we plot in black the characteristic starting at $x_p=1.25$ and indicate with thick lines the two disjoint pieces which contribute to $\Delta W(x_p)$.  (bottom) Above the horizontal axis we plot the height of the surface water $h(x,t)$ for a few values of $t$, and below the axis we show the final computed kick $\Delta W(x)$ as the shaded blue area. The water starts from a uniform height of $h_0$, then after some time we are left with a thin traveling wave moving downhill. An $x$ value whose characteristic has two disjoint pieces corresponds to a point where water infiltrates from $t=0$ until the height reaches 0, then at a later time a traveling wave deposits a second contribution of water as it crosses $x$.}
	\label{fig:characteristics}
\end{figure}

Characteristics $\hat{h}(y,y_\ell(x))$, computed using \eqref{eq:hhat0} for a few different values of $x$, are shown in Figure~\ref{fig:characteristics}(a). 
For the sample points $x\in[x_1^*,x_2^*]$, we note that there is a single interval $y\in[x,y_\ell(x)]$ where $\hat{h}\le h_0$, so that, upon evaluating the step function in \eqref{eq:Omegax0}, we obtain
\begin{equation}
	\label{eq:deltaw1}
	\Delta w(x)=\alpha \iota(x)\int_{x}^{y_\ell(x)}\frac{1}{\nu(y)}\ dy.
\end{equation}
In contrast, for sample points $x\in[0,\pi/2]\setminus[x_1^*,x_2^*]$, there are two intervals, $y\in[x,y_a]$ and $y\in[y_b,y_\ell(x)]$, where $\hat{h}\le h_0$, and thus
\begin{equation}
	\label{eq:deltaw2}
	\Delta w(x)=\alpha \iota(x)\Bigl( \int_{x}^{y_a}\frac{1}{\nu(y)}\ dy+\int_{y_b}^{y_\ell(x)}\frac{1}{\nu(y)} \ dy\Bigr).
\end{equation}
The first term represents the soil moisture that is from surface water originating close to $x$, whereas the second term represents water that traveled overland to $x$ from the bare region upslope of it, as described in Figure~\ref{fig:characteristics}. 


The bottom panel in Figure~\ref{fig:characteristics} shows some snapshots of the evolution of the surface water height $h(x,t)$. We see that the surface water initially covers the entire domain, then accumulates at the leading edge of the vegetation band, where the velocity is impeded by biomass. Eventually enough of the water infiltrates so that there is only a pulse of water traveling down the surface, which is eventually absorbed into the soil by $t\approx 8$. (In  dimensioned units this corresponds to almost an hour for the parameters of Table~\ref{tab:dim}.) 
Figure~\ref{fig:characteristics} also shows a plot of the soil moisture kick $\Delta w(x)$, $x\in[0,\pi]$, for this example. The maximum soil water depth is $\Delta w\approx 0.6$, 
and translates to a dimensioned depth of $\Delta W\approx 2.5 \ cm$ for parameters of Table~\ref{tab:dim}.
Note that while the biomass distribution is reflection symmetric about $x=\pi/2$, the resulting kick $\Delta w(x)$ is not. In this example, the soil moisture peak is slightly upslope of the vegetation peak. This is a necessary ingredient for the upslope migration of vegetation bands that we obtain in our model simulations. Observational studies of banded vegetation~\cite{deblauwe2012determinants} have detected a slow upslope colonization.

\section{Computation of stability thresholds for spatially uniform states}
\label{supp:transcrit}
Here we 
provide further details related to Figure \ref{fig:uniformthreshold}(a), which shows the threshold, in the $(T_d,H_0)$-plane, for stability of the zero biomass state with respect to spatially uniform perturbations.
Specifically, it shows three threshold curves, associated with different amounts of randomness, in a comparison to a curve for the case of periodic kicks, which has a threshold given by \eqref{eq:detthreshold}.
These curves are based on equation~\eqref{eq:ferandom-2}, derived in Section~\ref{sec:transcrit}, which gives the threshold condition as
\begin{equation}
	\mathbbm E\Bigl[\ln\Bigl(\frac{1+\zeta w_0}{1+\zeta\mu w_0} \Bigr)\Bigr]=\sigma\zeta \mathbbm E[\tau_d],
	\label{eq:ferandom-repeat}
\end{equation}
where $w_0$, the base state for the perturbation, is a random variable, as is $\mu=e^{-\sigma \tau_d}$ when the timing of kicks are random.
(Note that the conversion from non-dimensionalized to dimensionalized parameters, e.g. $\tau_d$ to $T_d$, is described in 
Section~\ref{sec:pulsedprecip:dim}, with further information in Table~\ref{tab:dim}.)

For the case of random kick strength only, we are able to derive, details below, a closed-form expression, in terms of a rapidly converging series, for the expected value on the left-hand-side of \eqref{eq:ferandom-repeat}.
For the other two random cases (random kick timing only and  fully random kick), we estimate this expected value numerically by averaging over a large number of flow kick cycles. We find that  $10^6$ cycles in the random kick strength only case is enough for the numerical estimate to match our analytical expression to 3 digits, so we use this number of cycles in the other two random cases. By comparing these computed expectations to the right-hand-side of \eqref{eq:ferandom-2} we  determine, via a bisection search, the threshold beyond which biomass perturbations grow.

\paragraph{Random kick strength only, $H_0\sim Exp(\lambda_H)$}

When the kicks are equally spaced at fixed intervals $\tau_d$, then  $w_0$ is the only random variable in~\eqref{eq:ferandom-repeat}. We determine its stationary distribution, denoted $f_W(w)$, from the flow-kick map given by $w_1=\mu w_0+\Delta w$. Here the kicks $\Delta w=\alpha h_0=\alpha H_0/{\cal H}_0$ are exponentially distributed, $\Delta w\sim Exp(\lambda_w)$ with rate parameter $\lambda_w=({\cal H}_0/\alpha)\lambda_H$; see Table~\ref{tab:dim}.
Further, if $w_1$ and $w_0$ follow the stationary distribution $f_W(w)$, then we can express $f_W(w)$ as a convolution
\begin{equation}
	\label{eq:conv}
	f_W(w)=(f_{\mu W}*f_{\Delta w})(w),
\end{equation}
where 
\begin{equation}
	\label{eq:scal}
	f_{\mu W}(x)=\frac{1}{\mu}f_W\Bigl(\frac{x}{\mu}\Bigr),\quad f_{\Delta W}(x)=\lambda_we^{-\lambda_w x},\quad x\ge 0.
\end{equation}
Combining \eqref{eq:conv} and \eqref{eq:scal}, and letting $x/\mu=y$, we obtain the following
integral equation for $f_W(w)$:
\begin{equation}
	\label{eq:fW:randomkick}
	f_W(w)=\lambda_we^{-\lambda_ww}\int_0^{w/\mu} f_W(y) \ e^{\lambda_w \mu y}\ dy,\quad w\ge 0.
\end{equation}

\begin{figure}
	\includegraphics[width=14cm]{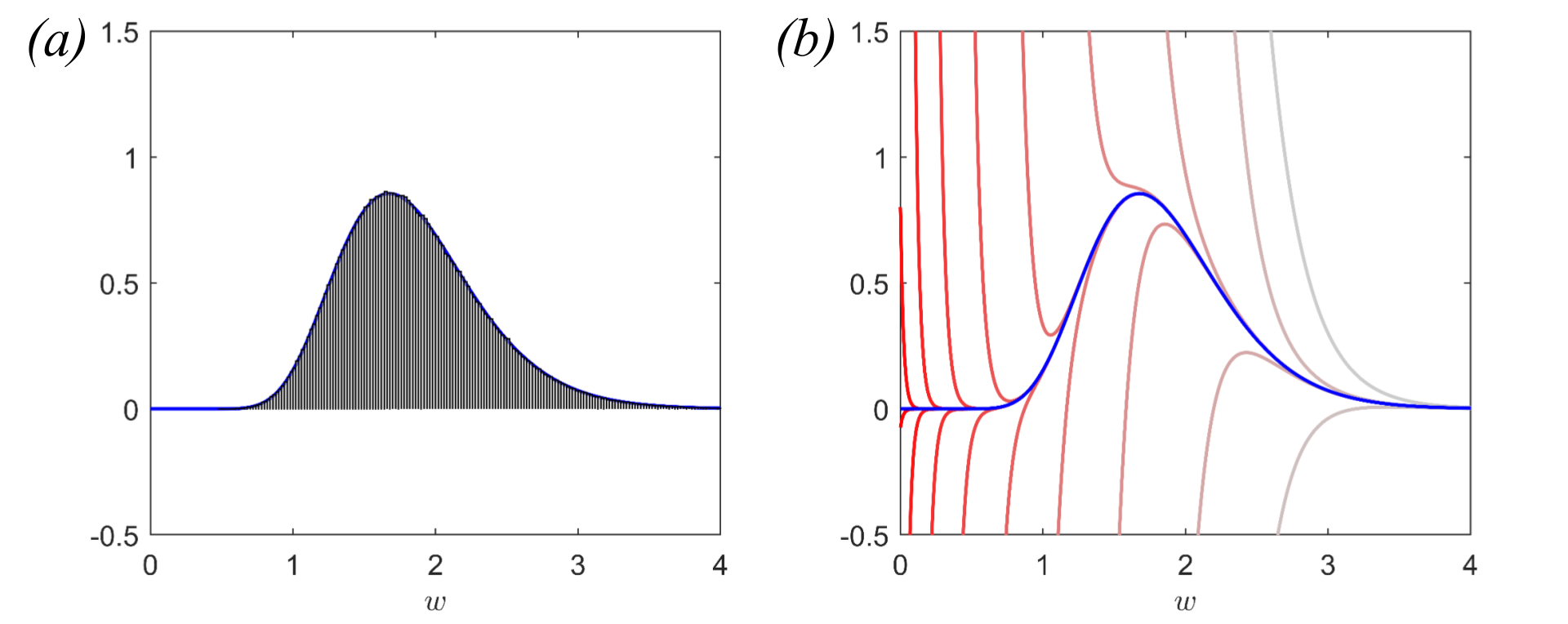}
	\centering
	\caption{(a) Comparison of \eqref{eq:series} with the distribution of $w$ estimated numerically using $10^6$ flow-kick iterations for $H_0$ = 1 cm and $T_d$ = 0.194 days, which is a point on the instability threshold of the zero-biomass state in Figure~\ref{fig:uniformthreshold}(a). (b) Comparison of \eqref{eq:series} (blue) and first 18 partial sums (gray to red). }
	\label{fig:distributions}
\end{figure}

It follows from~\eqref{eq:fW:randomkick} that the distribution $f_W(w)$ is a fixed point of the map ${\cal T}$ given by
\begin{equation*}
	{\cal T}(f(w))=\lambda_we^{-\lambda_ww}\int_0^{w/\mu} f(y) \ e^{\lambda_w \mu y}\ dy.
\end{equation*}
Iterating this map $n$ times, from the initial exponential distribution $f^0_W(w)=\lambda_w e^{-\lambda_w w}$, and taking the limit $n\to \infty$, suggests the following series ansatz for $f_W(w)$:
\begin{equation}
	\label{eq:series}
	f_W(w)=\lambda_w\sum_{k=0}^\infty A_k(\mu)\ e^{-\lambda_ww/\mu^k},\quad w\ge 0.
\end{equation}
If we 
equate coefficients of each term $e^{-\lambda_w w/\mu^k}$, $k\ge 1$, on the left- and right-hand-sides of \eqref{eq:fW:randomkick}, we obtain
the recurrence relation
\begin{equation}
	\label{eq:akrecurrence}
	A_k(\mu)= -\Bigl(\frac{\mu^{k-1}}{1-\mu^k}\Bigr)A_{k-1}(\mu),\quad k\ge 1.
\end{equation}
It follows  that
\begin{equation}
	\label{eq:Akequation}
	A_k(\mu)=(-1)^k 
	A_0(\mu)\prod_{n=1}^{k}
	\Bigl(\frac{\mu^{n-1}}{1-\mu^n}
	\Bigr)\  ,\quad k\ge 1.
\end{equation}
Moreover, if we equate the coefficients of the $e^{-\lambda_w w}$ terms in \eqref{eq:fW:randomkick}, we determine that
\begin{equation}
	\label{eq:A0_relation}
	A_0(\mu)=-\sum_{k=0}^\infty A_k(\mu)\ \Bigl(\frac{\mu^k}{1-\mu^{k+1}}\Bigr).
\end{equation}
Dividing through by $A_0(\mu)$, and combining with \eqref{eq:Akequation}, we arrive at the following identity, which is proved below,
\begin{equation}
	\label{eq:PochIndentity}
	\sum_{k=1}^\infty (-1)^{k+1}\prod_{n=1}^{k} \Bigl(\frac{\mu^{n-1}}{1-\mu^n}\Bigr)=1.
\end{equation}
Finally, $A_0(\mu)$ is determined by normalization, $\int_0^\infty f_W(w) \ dw=1.$ Figure~\ref{fig:distributions} shows convergence of the partial sums of this representation of $f_W$, as well as agreement with the distribution of $w$ generated directly from  a large number of flow-kick cycles.

We use \eqref{eq:series} to evaluate the left-hand-side of \eqref{eq:ferandom-repeat} and find that
\begin{equation}
	\label{eq:expintexpectation}
	\mathbbm E\Bigl[\ln\Bigl(\frac{1+\zeta w_0}{1+\zeta\mu w_0} \Bigr)\Bigr]
	=\sum_{k=0}^\infty A_k(\mu)\mu^k\Bigl(
	E_1\Bigl(\frac{\lambda_w}{\zeta\mu^k}\Bigr)e^{\lambda_w/\zeta \mu^k}- E_1\Bigl(\frac{\lambda_w}{\zeta\mu^{k+1}}\Bigr)e^{\lambda_w/\zeta \mu^{k+1}}
	\Bigr),
\end{equation}
where $E_1(z)=\int_z^\infty \frac{1}{y} e^{-y} \ dy$ is the exponential integral. 
For small values of $z=\lambda_w/\zeta\mu^k$ in \eqref{eq:expintexpectation}, we can accurately evaluate the products $E_1(z)e^z$,  but as z becomes large, the $E_1(z)$ factor becomes small while the $e^z$ factor becomes large, resulting in numerical error. When $z>1$, we find success using Lentz's algorithm with the continued fraction expansion:
\begin{equation*}
	E_1(z)e^z = \frac{1}{z+\frac{1}{1+\frac{1}{z+\frac{2}{1+\frac{2}{z+\frac{3}{...}}}}}}
\end{equation*}
Specifically, Lentz's algorithm computes truncations of the continued fraction, and stops when the difference between successive truncations is small. We find a tolerance of 1e-10 to work well.

\paragraph{Proof of the identify~\eqref{eq:PochIndentity}} Our proof makes use of the q-Pochhammer symbol~\cite{gasper2011basic}, defined by
\begin{equation*}
	(p,q)_k = \prod_{n=0}^{k-1}(1-pq^n), \quad k\ge 1,
\end{equation*}
with $(p,q)_0=1$, 
as well as the series representation
\begin{equation*}
	(x,q)_{\infty} = \sum_{k=0}^{\infty}(-1)^{k}\frac{q^{k(k-1)/2}}{(q;q)_k}x^k, \quad |q|<1.
\end{equation*}
Specifically, we find that the left-hand-side of \eqref{eq:PochIndentity} is
\begin{align*}
	\sum_{k=1}^{\infty} (-1)^{k+1} \prod_{n=1}^k   \frac{\mu^{n-1}}{1-\mu^n} &= \sum_{k=1}^{\infty} (-1)^{k+1} \frac{\mu^{k(k-1)/2}}{(\mu;\mu)_k} \\ &=
	1-\sum_{k=0}^{\infty} (-1)^{k} \frac{\mu^{k(k-1)/2}}{(\mu;\mu)_k} \\ &=
	1-(1;\mu)_\infty \\ &= 1  \quad \square
\end{align*}
\bibliographystyle{unsrt}
\bibliography{references}
\end{document}